\documentclass[useAMS,usenatbib,usegraphicx,pdftex]{mn2e}
\usepackage{url, times}
\usepackage{amsmath}
\usepackage{cases}
\allowdisplaybreaks

\title[Blazhko RR\,Lyrae light curves as modulated signals]
{Blazhko RR\,Lyrae light curves as modulated signals}
\author[Benk\H{o} et al.]{
J.~M. Benk\H{o}\thanks{E-mail: benko@konkoly.hu},
R. Szab\'o, and
M. Papar\'o\\
Konkoly Observatory, Konkoly Thege M. u. 15-17., H-1121 Budapest, Hungary}
\begin{document}

\date{Accepted 2011 June 23. Received 2011 June 23; in original form 2011 April 28}

\pagerange{\pageref{firstpage}--\pageref{lastpage}} \pubyear{2011}

\maketitle

\label{firstpage}

\begin{abstract}
We present an analytical formalism for the description of Blazhko 
RR\,Lyrae light curves in which employ a treatment for the 
amplitude and frequency modulations in a manner similar to the 
theory of electronic signal transmitting. We assume 
monoperiodic RR Lyrae light curves as carrier waves and modulate their 
amplitude (AM), frequency (FM), phase (PM), and as a 
general case we discuss simultaneous AM and FM. 
The main advantages of this handling are the following: 
(i) The mathematical formalism naturally explains numerous 
light curve characteristics found in Blazhko 
RR\,Lyrae stars such as mean 
brightness variations, complicated envelope curves, 
non-sinusoidal frequency variations. 
(ii) Our description also explains
properties of the Fourier spectra such as apparent higher-order 
multiplets, amplitude distribution of the side peaks, 
the appearance of the modulation frequency itself and
its harmonics. In addition, comparing to the traditional method, our 
light curve solutions reduce the number of necessary parameters. 
This formalism can be applied to any type of modulated 
light curves, not just for Blazhko RR\,Lyrae stars.
\end{abstract}


\section{Introduction}

The Blazhko effect \citep{Bla07} is a 
periodic amplitude and phase variation of the 
RR\,Lyrae variable stars' light curve.
The typical cycle lengths of these variations are
about of 10-100 times longer than the main pulsation periods 
(0.3$-$0.7~d). 
Almost half of the RR\,Lyrae stars pulsating in their 
fundamental mode (type RRab) and a smaller but non-negligible fraction of 
the first overtone mode pulsating stars (type RRc) show the effect 
\citep{Jur09, Cha09, Kol10, Ben10}.	
It is usually interpreted as a modulation or a beating phenomenon, 
but both hypotheses have their own problems.
The beating picture describes the main feature of the
light curves and Fourier spectra well (see \citealt{Bre06, Kol06}), 
but reproducing phase variations, multiplet structures found in certain stars' 
Fourier spectra \citep{JurMW, Cha10} in this framework is not possible.
On the other side, the stars showing doublet structures in their Fourier 
spectra \citep{Alc00, Alc03, MP03} seemed to be contradicted
with the modulation picture. 

In this paper  we describe the Blazhko effect  
as a modulation, and derive the mathematical consequences of
this assumption by developing a consistent analytical framework.
Using this framework we demonstrate that many light curve 
characteristics are naturally identified as mathematical
consequences of the modulation assumption. 
By disentangling these features we get closer to the 
physics behind the Blazhko effect. 

The possibility of the modulation/Blazhko effect have been raised 
for many types of pulsating stars from Cepheids to $\delta$~Scuti 
stars (see e.g. \citealt{Koe01, Hen05, Mos09, Bre10, Por11}). 
The main motivation of this paper is the mathematical description of
the Blazhko RR\,Lyrae stars' light curves and investigate 
their properties. Most of our results can be applied 
directly to any other types of variable stars, where modulation is suspected. 
Our deduced formulae and the related phenomena may help to prove 
or reject the modulation hypothesis.    

The basic idea of this paper was raised in \citet{Ben09}.
Modulation is a technique that has been 
used in electronic communication for a long time, 
mostly for transmitting information signal via a radio carrier wave.
In those cases, the carrier wave is a sinusoidal electromagnetic (radio) wave
that is modulated by a (generally non-periodic) information signal (e.g. speech, music).
In this paper the formalism developed by engineers 
for broadcasting radio signals
has been modified such a way that we assumed a monoperiodic non-modulated 
RR\,Lyrae light variation as a carrier wave.
While for communication usually only one type of modulation is applied
we allow both types of modulations (amplitude and angle). 
 
In Section~\ref{Basic} 
we present a collection of classical formulae that
are well-known in physics of telecommunications 
\citep{Car22, vdP30, Rod31}.
Some of the more complicated cases 
(multiple modulations, recursive or cascade modulations)
were investigated by mathematicians who developed the theory of
electric sound synthesizers in the years of 1960s and '70s.   
The formulae are modified to describe the modulated light curves
 in Sec.~\ref{Sec_Bl} and investigated by a 
step-by-step process from the simplest cases to the more complex ones. 
Section 4 compares the numerical behaviour of 
the traditional method and this one. 
Section 5 summarizes our results.

\section{Basic formulae}\label{Basic}

In this section we briefly review some classical 
definitions and formulae (see e.g. \citealt{NK04, Sch03})
that will be used through  the next sections. 
The simplest periodic signal is a sinusoidal function.
It has three basic parameters: 
amplitude, frequency and phase and any of these can be 
modulated. 

\subsection{Amplitude modulation}\label{Basic_AM}
 
The amplitude modulation (AM) is the simplest of the three cases.
Let the carrier wave $c(t)$ be a simple sinusoidal signal as 
\begin{equation}\label{car}
c(t)=U_{\mathrm c}\sin(2\pi f_{\mathrm c} t + \varphi_{\mathrm c}),
\end{equation}  
where $U_{\mathrm c}$, $f_{\mathrm c}$ and $\varphi_{\mathrm c}$ 
constant parameters are the amplitude, frequency and initial phase
of the carrier wave, respectively.

Let $U_{\mathrm m}(t)$ represent an arbitrary waveform that 
is the message to be transmitted.  
The transmitter uses the information signal $U_{\mathrm m}(t)$ to vary the 
amplitude of the carrier $U_{\mathrm c}$ to produce a modulated signal: 
\begin{equation}\label{def_am}
  U_{\mathrm{AM}}(t)=\left[ U_{\mathrm c}+
U_{\mathrm m} \left( t \right) \right] 
\sin(2\pi f_{\mathrm c} t + \varphi_{\mathrm c}).
\end{equation}

In the simplest case, where the modulation is also sinusoidal
\begin{equation}\label{sin_AM}
U_{\mathrm m}(t)
=U^{\mathrm A}_{\mathrm m}\sin(2\pi f_{\mathrm m} t + 
\varphi^{\mathrm A}_{\mathrm m}).
\end{equation} 
Substituting Eq.~(\ref{sin_AM}) into (\ref{def_am}) and 
using basic trigonometrical identities, expression 
(\ref{def_am}) can be rewritten as
\begin{multline}\label{AM1}
U_{\mathrm{AM}}(t)=U_{\mathrm c}\sin (2\pi f_{\mathrm c}t + \varphi_{\mathrm c}) + {} \\ 
\frac{U^{\mathrm A}_{\mathrm m}}{2}\left\{ 
\sin \left[ 2\pi \left( f_{\mathrm c} - f_{\mathrm m} \right) t 
+\varphi^{-} \right] +
\sin \left[ 2\pi \left( f_{\mathrm c} + f_{\mathrm m} \right) t 
+\varphi^{+} \right] 
\right\},
\end{multline}
where 
$\varphi^{-}=\varphi_{\mathrm c}-\varphi^{\mathrm A}_{\mathrm m}+\pi /2$ 
and 
$\varphi^{+}=\varphi_{\mathrm c}+\varphi^{\mathrm A}_{\mathrm m}-\pi /2$.
The initial shifts ($\pm \pi/2$) appear because we used 
(through all this paper) a sinusoidal representation instead of 
sin and cos functions.

The exact analytical Fourier transformation of (\ref{AM1}) 
is given in Appendix \ref{AAF},
however, the basic structure of the frequency spectrum can be easily 
read off from Eq.~(\ref{AM1}). 
Since the Fourier spectrum of a single sinusoidal function shows a 
peak at the frequency of the sinusoid, from the above expression 
(\ref{AM1}) the well-known triplet structure
composed of the peaks 
$f_{\mathrm c}$ and $f_{\mathrm c}\pm f_{\mathrm m}$ can be seen. 
The amplitude of the side-peaks $f_{\mathrm c}\pm f_{\mathrm m}$ are 
always equal. The Fourier amplitude of the carrier wave 
($\pi \sqrt{2\pi} U_{\mathrm c}$), 
that represents the energy at the carrier frequency, is constant.

The ratio of the carrier wave amplitude $A(f_{\mathrm c})$ and 
the side peaks $A(f_{\mathrm c}\pm f_{\mathrm m})$ 
are  connected to the modulation depth. 
We rewrite Eq.~(\ref{def_am}) as
\begin{equation}\label{mod_d}
  U_{\mathrm{AM}}(t)=\left[ 1 + \frac{U_{\mathrm m} \left( t \right)}{U_{\mathrm c}} \right] c(t).
\end{equation}
If $U_{\mathrm m}(t)$ is a bounded function, 
let $U_{\mathrm m}^{\mathrm{max}}$ represent the maximum value
of this modulation function, then {\it modulation depth} 
is defined as $h=U_{\mathrm m}^{\mathrm{max}}/U_{\mathrm c}$.
In the above discussed sinusoidal case 
$h=U^{\mathrm A}_{\mathrm m}/U_{\mathrm c}$ and 
$A(f_{\mathrm c}\pm f_{\mathrm m})/A(f_{\mathrm c})= \frac{1}{2}h$.
In other words, 
the amplitude of the central peak is twice of the side peak highs.

\subsection{Angle modulations}\label{Basic_FM}

The phase and frequency modulations together are called 
angle modulations. Since, when we assume the sinusoidal carrier 
wave Eq.~(\ref{car}) as $c(t)=U_{\mathrm c}\sin[\Theta(t)]$, 
the $\Theta(t)=2\pi f_{\mathrm c} t + \varphi_{\mathrm c}$ denotes the 
angle part of the function.

Phase modulation (PM) changes the phase angle of the carrier signal.
Suppose that the modulating or message signal is $U_{\mathrm m}(t)$,
then  $\Theta(t)=2\pi f_{\mathrm c} t + [\varphi_{\mathrm c}
+ U_{\mathrm m}(t)]$.
Let $U_{\mathrm m}(t)$ be again a bounded function. In this case
we can define a constant as: 
$k_{\mathrm{PM}}=\vert U^{\mathrm {max}}_{\mathrm m}(t)\vert/2$.
This transforms the modulated signal
\begin{equation}\label{def_pm}
 U_{\mathrm{PM}}(t)=U_{\mathrm c} \sin \left[2\pi f_{\mathrm c} t 
+ k_{\mathrm{PM}} U^{\mathrm P}_{\mathrm m} \left(t \right) 
 + \varphi_{\mathrm c} \right],
\end{equation}
where $\vert U^{\mathrm P}_{\mathrm m}(t) \vert \le 1$.  
The {\it instantaneous frequency} of the 
modulated signal is 
\begin{equation}
f(t)=\frac{{\mathrm d} \Theta}{{\mathrm d} t}=f_{\mathrm c}
+k_{\mathrm{PM}}\frac{{\mathrm d} U^{\mathrm P}_{\mathrm m}(t)}{{\mathrm d} t}.
\end{equation}

Frequency modulation (FM) uses the modulation 
signal $U_{\mathrm m}(t)$ to vary the carrier frequency.
$\Theta(t)=2\pi f(t) t + \varphi_{\mathrm c}$ and here the instantaneous
frequency $f(t)$ is modulated by the signal of 
$k_{\mathrm{FM}}U^{\mathrm F}_{\mathrm m}(t)$ as 
\begin{equation}\label{FMi}
f(t)=f_{\mathrm c}+k_{\mathrm{FM}}U^{\mathrm F}_{\mathrm m}(t).
\end{equation}
In this equation $k_{\mathrm{FM}}$ is the frequency deviation, 
which represents the maximum shift from $f_{\mathrm c}$ in one direction,
assuming $U^{\mathrm F}_{\mathrm m}(t)$ is limited to the 
range ($-1,\dots,+1$).
Using the definitions of the instantaneous frequency and phase, 
expression (\ref{FMi}) can be rewritten as 
$\Theta(t)=2\pi f_{\mathrm c} t 
+ 2\pi k_{\mathrm{FM}} \int_0^t 
U^{\mathrm F}_{\mathrm m}(\tau) d\tau+\varphi_{\mathrm c}$.
The modulated signal is
\begin{equation}\label{def_fm}
 U_{\mathrm{FM}}(t)=
U_{\mathrm c} \sin \left[ 2\pi f_{\mathrm c} t + 2\pi  k_{\mathrm{FM}} 
\int_0^t U^{\mathrm F}_{\mathrm m}\left( \tau \right) d\tau 
+  \varphi_{\mathrm c} \right].
\end{equation}
This definition of FM is the least intuitive of the three
Eqs~(\ref{def_am}, \ref{def_pm} and \ref{def_fm}). 
If we compare Eqs~(\ref{def_pm}) and (\ref{def_fm})
we realize that the modulation signals are in derivative-integral 
connection with each other. In practice, we have
modulating signals that can be represented by analytical functions.
Therefore, when we detect an FM or PM signal without any 
previous knowledge about them, it is  impossible
to distinguish between FM and PM signals.

\begin{figure} 
\includegraphics[angle=270, width=8.5cm]{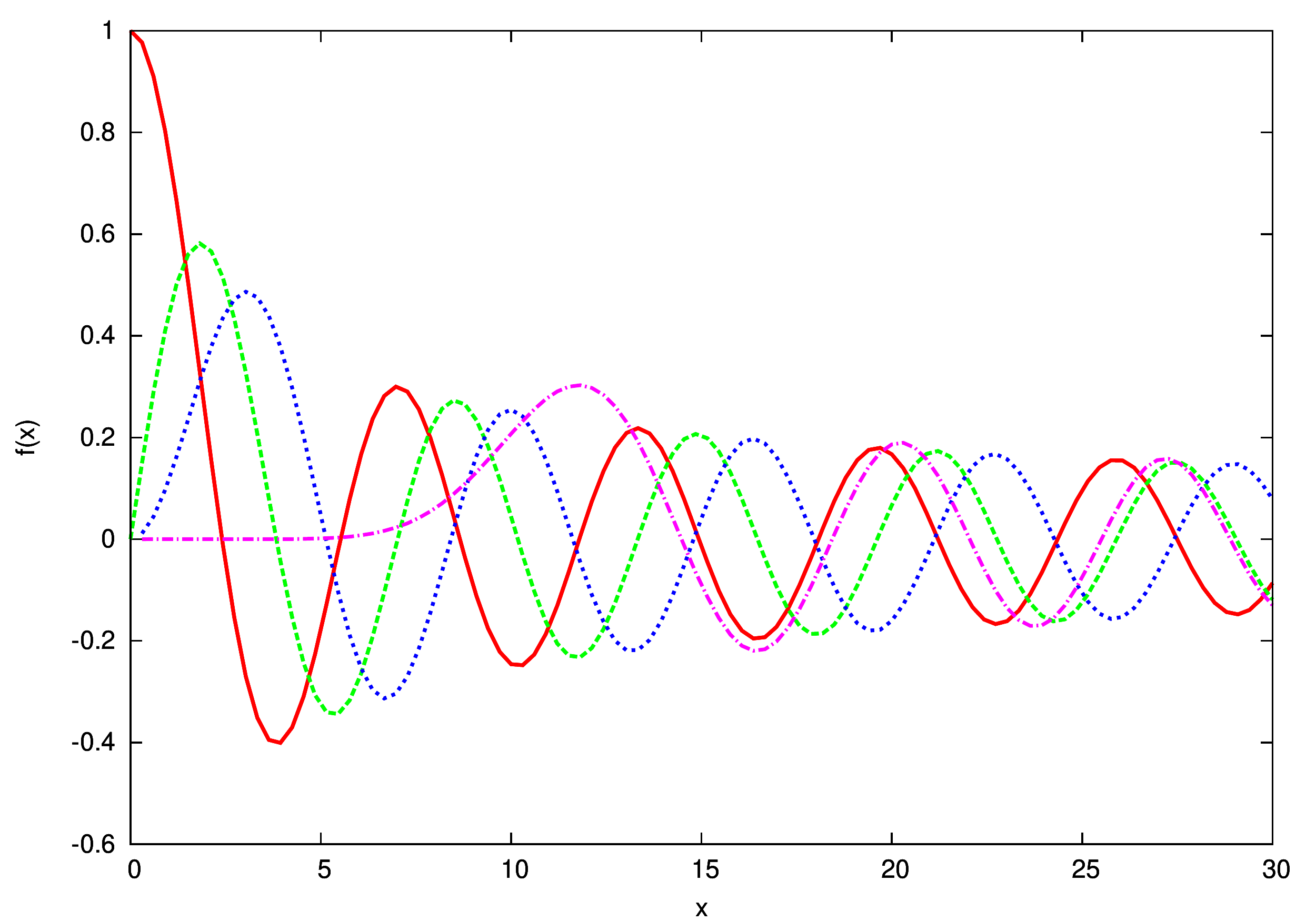}
\caption{
The graph of the first three and the 10th 
Bessel functions of first kind
with integer order $k$ for the value of $x$.
$J_0(x)$ -- (red) continuous line, $J_1(x)$ -- (green) dashed line,
$J_2(x)$ -- (blue) dotted line, 
$J_{10}(x)$ -- (purple) dash-dotted line.
}\label{BesselJ}
\end{figure}

First, let the modulating signal be
represented by a sinusoidal wave with a frequency $f_{\mathrm m}$. 
The integral of such a signal is
\begin{equation}
U^{\mathrm F}_{\mathrm m}(t)=
\frac{U^{\mathrm F}_{\mathrm m}}{2\pi f_{\mathrm m}} \sin \left( 2\pi 
f_{\mathrm m} t + \varphi^{\mathrm F}_{\mathrm m} \right).
\end{equation}
Thus, in this case Eq.~(\ref{def_fm}) gives
\begin{equation}\label{mod_FM}
U_{\mathrm{FM}}(t)=
U_{\mathrm c} \sin \left[ 2\pi f_{\mathrm c} t + \eta
\sin \left( 2\pi f_{\mathrm m} t + 
\varphi^{\mathrm F}_{\mathrm m} \right) +  
\varphi_{\mathrm c} \right],
\end{equation}
where the {\it modulation index} is defined as 
$\eta= (k_{\mathrm{FM}} U^{\mathrm F}_{\mathrm m})/f_{\mathrm m}$.
Eq.~(\ref{mod_FM}) can be deduced from Eq.~(\ref{def_pm}) as well.
The only difference is the value of 
$\eta=k_{\mathrm{PM}} U^{\mathrm P}_{\mathrm m}$, 
which is independent of the modulation frequency $f_{\mathrm m}$. 
Let us transcribe Eq.~(\ref{mod_FM}) using relations for
trigonometrical and Bessel functions \citep{AS72} as:
\begin{equation}\label{mod_Chowning}
U_{\mathrm{FM}}(t)=U_{\mathrm c} \sum^{\infty}_{k=-\infty} J_k(\eta) \sin
\left[ 2 \pi \left( f_{\mathrm c} + 
kf_{\mathrm m} \right) t + k\varphi_{\mathrm m} 
+ \varphi_{\mathrm c} \right],
\end{equation}
where $J_k(\eta)$ is the Bessel function of first kind 
with integer order $k$ for the value of $\eta$ 
(Fig.~\ref{BesselJ}); $\varphi_{\mathrm m}$
denotes either $\varphi^{\mathrm F}_{\mathrm m}$ 
or $\varphi^{\mathrm P}_{\mathrm m}$. 
This formula is known as the Chowning relation \citep{Ch73}.
Although it had been deduced by many different authors formerly, 
Chowning recognized its key role 
in the electronic sound creation method called FM synthesis.

Similarly to Eq.~(\ref{AM1}) expression 
(\ref{mod_Chowning}) also helps to imagine the
spectrum. It is made up of a carrier at $f_{\mathrm c}$ and symmetrically 
placed side peaks separated by $f_{\mathrm m}$. 
The amplitudes follow the Bessel functions.  
The behavior of the Bessel functions is well known: except for small 
arguments ($x < k$), they behave like damped sine functions 
(see also, Fig.~\ref{BesselJ}). For higher indices the higher
order side peaks gradually become more and more important.
As a consequence, the amplitude of the central peak gets reduced.
The frequency spectrum of an actual FM signal has 
an infinite number of side peak 
components, although they become negligibly small beyond a 
point.

If $\vert \eta \vert \ll 1$, we find that $J_0(\eta)\approx 1$, 
$\vert J_{\pm1} \vert = \eta /2$ and $J_k \approx 0$  for $k>1$.
That is, the spectrum can be approximated with an equidistant 
triplet similarly to AM, but
the character of the signal differs from AM: the total amplitude 
of the modulated wave remains constant. 
When $\eta$ increases the amplitude of
side peaks also increases, but the Fourier 
amplitude of the carrier decreases.
In other words, the side peaks could be larger than the 
central peak, on the other hand higher order side frequencies 
could also be of larger amplitude than the lower order ones.

A more general case is formulated by \cite{Sch77}
\begin{multline}\label{mod_Sch}
U_{\mathrm{FM}}(t)=
U_{\mathrm c} \sin \left[ 2\pi f_{\mathrm c} t + 
\sum_{p=1}^q U_{\mathrm m}^{(p)} \sin \left( 2\pi f^{(p)}_{\mathrm m} t 
+ \varphi^{(p)}_{\mathrm m} \right) 
+  \varphi_{\mathrm c} \right] {} \\
= U_{\mathrm c} 
\sum_{k_p= -\infty}^{\infty} \dots \sum_{k_1= -\infty}^{\infty}
\left[ \prod_{p=1}^q J_{k_p}(U_{\mathrm m}^{(p)}) \right]
\sin \Bigg[  2\pi f_{\mathrm c} t + {} \\
\sum_{p=1}^q k_p 
\left( 2\pi f^{(p)}_{\mathrm m} t + 
\varphi^{(p)}_{\mathrm m} \right) + \varphi_{\mathrm c} \Bigg].
\end{multline}
Here the modulating signal is assumed to be a linear combination of 
a finite number of sinusoidal functions with arbitrary frequencies
$f^{(p)}_{\mathrm m}$, amplitudes $U_{\mathrm m}^{(p)}$ and 
phases $\varphi^{(p)}_{\mathrm m}$, $p=1, 2, \dots, q$. 
The spectrum contains equidistant frequencies on both 
sides of the carrier frequency. The amplitudes of the side peaks 
$f_{\mathrm c}\pm k_p f^{(p)}_{\mathrm m}$ 
are determined by products of Bessel functions. 

\subsection{Combined modulation}\label{Basic_Bela}

In practice, the electronic circuits that generate 
modulated signals generally produce a mixture of 
amplitude and angle modulations. 
This combined modulation is disliked in radio 
techniques but welcomed in sound synthesis and as we will see
they appear in the case of Blazhko RR\,Lyrae stars, as well. 
Let us overview the basic 
phenomena of combined modulations following \cite{Car66}.
We start with the simplest case: both the 
AM and FM are sinusoidal and their frequencies are 
the same. 
\begin{multline}\label{def_comb}
  U_{\mathrm{Comb}}(t)=U_{\mathrm c} 
\left( 1 + h \sin  2 \pi f_{\mathrm m} t \right)  
\sin\left[ 2\pi f_{\mathrm c} t + \right. {} \\
\left. \eta \sin \left( 2\pi f_{\mathrm m} t 
+ \phi_{\mathrm m}\right) +\varphi_{\mathrm c} \right].
\end{multline} 
By suitable choice of the starting epoch, without any restriction of the
general validity we can set $\varphi_{\mathrm c}= 0$. 
Here $\phi_{\mathrm m}$ is the
relative phase difference between the modulating 
FM and AM signals. Other designations are the same as before. 
The third term of the  product (\ref{def_comb})
is the same as in Eq.~(\ref{mod_FM}), therefore, after 
applying the Chowning relation (\ref{mod_Chowning})  
\begin{multline}
 U_{\mathrm{Comb}}(t)= U_{\mathrm c} \left( 1 + 
h \sin  2 \pi f_{\mathrm m} t \right)\cdot \\
\sum^{\infty}_{k=-\infty} J_k(\eta) \sin
\left[ 2 \pi \left( f_{\mathrm c} +
kf_{\mathrm m} \right) t + k\phi_{\mathrm m} \right].
\end{multline}
This expression results in an infinite number of amplitude
modulated waves. After trigonometrical transformations
we get:
\begin{multline}\label{mod_Comb}
 U_{\mathrm{Comb}}(t)= U_{\mathrm c}
\sum^{\infty}_{k=-\infty} \bigg\{ J_k(\eta) 
\sin \left[ 2 \pi \left( f_{\mathrm c} + 
kf_{\mathrm m} \right) t + k\phi_{\mathrm m} \right] + {} \\
\left. \frac{h}{2}J_{k-1}(\eta ) 
\sin \left[ 2 \pi \left( f_{\mathrm c} + kf_{\mathrm m} \right) t + 
\left( k-1 \right) \phi_{\mathrm m} - \frac{\pi}{2} \right] + \right. {} \\ 
\left. \frac{h}{2}J_{k+1}(\eta ) 
\sin \left[ 2 \pi \left( f_{\mathrm c} + kf_{\mathrm m} \right) t + 
\left( k+1 \right) \phi_{\mathrm m} + \frac{\pi}{2} \right] \right\}.
\end{multline}
It can be seen that each terms consists of three sinusoidal
functions with different phases. On the basis of expression 
(\ref{mod_Comb}), the spectrum of combined modulation (\ref{def_comb}) is comprehensible 
as a combination of three FM spectra. The peaks are 
at the same places as the frequencies
of the spectrum of (\ref{mod_Chowning}), but the amplitudes of a pair of side peaks
are generally asymmetrical. 
Using some trigonometrical identities, the
rules of summation of parallel harmonic oscillations and relations for Bessel functions 
we arrive to the expression for the Fourier amplitudes of a certain frequency:
\begin{multline}\label{Camp}
A(f_{\mathrm c} + k f_{\mathrm m}) \sim 
U_{\mathrm c} \left\{ J^2_k(\eta) \left( 1 - 
\frac{h k}{\eta} \sin \phi_{\mathrm m} \right)^2 + \right .\\
\left. \frac{h^2}{4} \cos^2 \phi_{\mathrm m} 
\left[ J_{k+1}(\eta)-J_{k-1}(\eta) \right]^2 \right\}^{\frac{1}{2}},
\end{multline}
($k=0, \pm 1, \pm 2, \dots$).
Introducing the {\it power difference of the side peaks} 
as it was done by \cite{Sz09}
$\Delta_l:=A^2(f_{\mathrm c} + l f_{\mathrm m})- A^2(f_{\mathrm c} - l f_{\mathrm m})$,
where $l=1, 2, 3 \dots$ and taking into account formula (\ref{Camp}),
we get 
\begin{equation}\label{asymm}
\Delta_l=-4\frac{h l}{\eta} U_{\mathrm c}^2J^2_l(\eta) \sin \phi_{\mathrm m}.
\end{equation}
This formula is a direct generalisation of the formulae given by \cite{Sz09}
for $l=1$ and $l=2$.
It is evident, that this asymmetry parameter depends 
only on $\phi_{\mathrm m}$, the relative phase of AM and FM. 
The left hand side peaks are higher than the right hand side ones ($\Delta_l<0$), 
if $0 < \phi_{\mathrm m} < \pi$, otherwise the situation is opposite: 
$\pi < \phi_{\mathrm m} < 2\pi$ and $\Delta_l>0$. In those very special 
cases, where $\phi_{\mathrm m}=0$ or $\phi_{\mathrm m}=\pi$ the side peaks'
amplitudes are equal. We have to note that 
if one of the modulations from AM and FM  dominates in the combined case
($h \ll \eta$ or $\eta \ll 1$),
the amplitude of the side peaks are almost the same.

\section{Blazhko modulation}\label{Sec_Bl}

RR\,Lyrae light curves traditionally 
are described by a Fourier series of a
limited number of terms. In Blazhko modulated RR\,Lyrae stars the
sum builds up from terms of harmonics of the main pulsation frequency,
side peaks due to the modulation and the modulation frequency and even
its harmonics:
\begin{equation}\label{old}
m(t)=A_{0}+ 
\sum_{i=1}^N{A_{i} \sin 
\left[ 2\pi F_i t + \Phi_i \right]
},
\end{equation}
where either $F_i=j f_0$, ($j=1,2,\dots, n$); or
$F_i= k f_{\mathrm m}$, ($k=1,2,\dots, m$); or 
$F_i= j' f_0 + k' f_{\mathrm m}$, 
($j'=1,2,\dots, n'$, $k'=1,2,\dots, m'$);
$F_i= j'' f_0 - k'' f_{\mathrm m}$, 
($j''=1,2,\dots, n''$, $k''=1,2,\dots, m''$);
$f_0$ and $f_{\mathrm m}$ are the main pulsation frequency and
the modulation one, respectively.
The amplitudes $A_i$ and phases $\Phi_i$ are considered
as independent quantities and determined  by a non-linear
fit. The necessary number of 
parameters for a complete light curve solution is $2N+3$ 
(amplitudes and phases and two frequencies and the zero point $A_{0}$).
The number of parameters can be as high as 500-600 for 
a long time series of good quality (see e.g. \citealt{Cha10}). 

In the next subsections we show how the modulation
paradigm can be applied and what advantages it has compared
to this traditional handling Eq.~(\ref{old}).

\subsection{Blazhko stars with AM}\label{Sec_AM}

To start with, we discuss Blazhko stars' light curves with pure AM effect,
although the recent space-born data suggest 
that all Blazhko RR\,Lyrae stars show 
amplitude modulation and simultaneous period changes \citep{Cha10, Ben10, Por10}.
We follow a step-by-step generalization process that allows us
to separate effects more clearly. 
We note that the most striking feature of a Blazhko 
RR\,Lyrae light curve is the amplitude variation, which
is generally easy to find and in many cases the only detectable modulation
(see \citealt{Sto10} and references therein).

\begin{figure} 
\includegraphics[width=8.5cm]{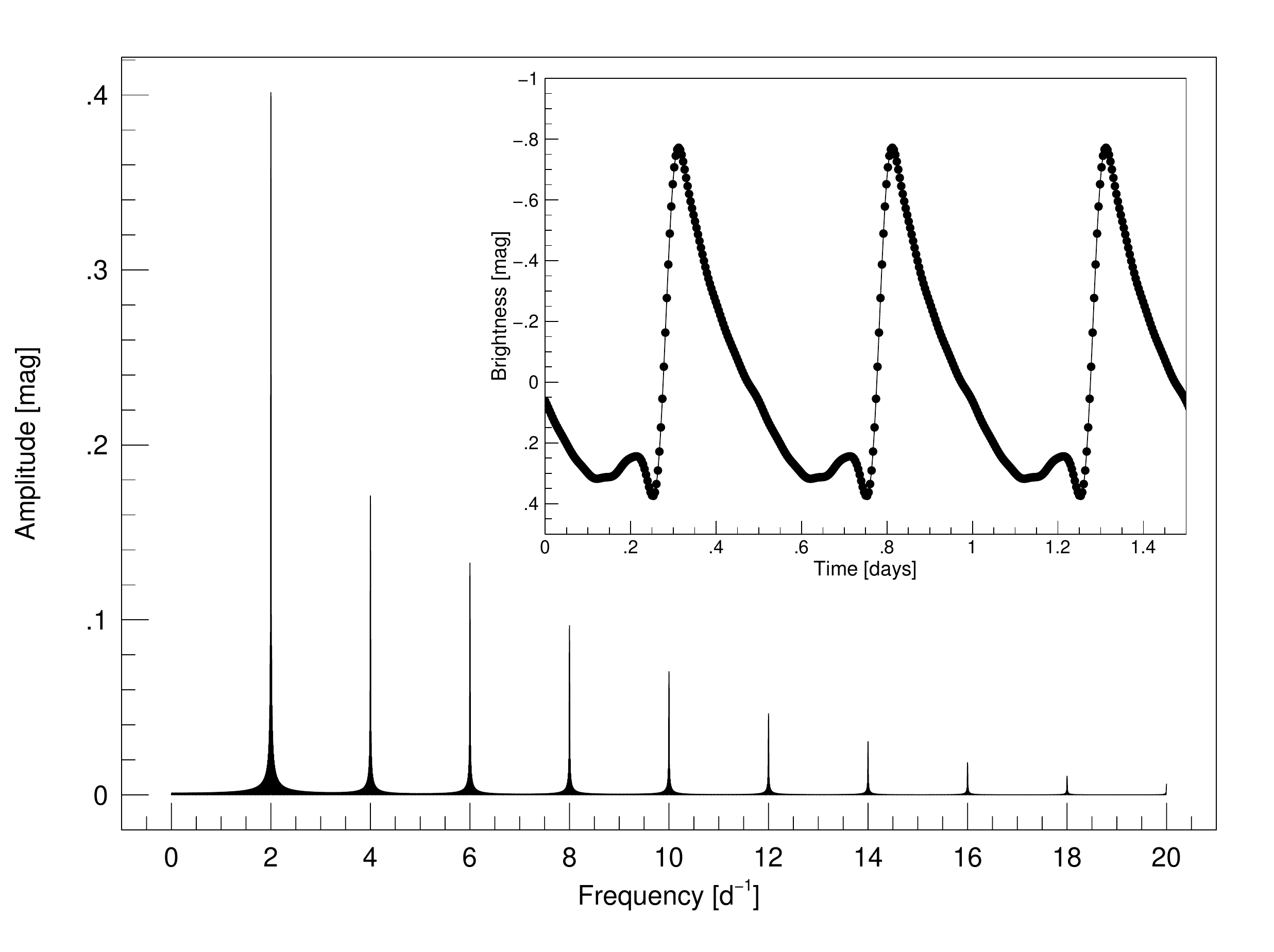}
\caption{Fourier amplitude spectrum of the 
artificial RR\,Lyrae light curve
acting as a carrier wave of all modulated 
light curves constructed in this paper (main panel), 
and a part of the light curve itself (insert). 
}\label{carrier}
\end{figure}

To apply the framework described in Sec.~\ref{Basic_AM} to an RR\,Lyrae 
light curve the course-book formulae need some extensions. 
We choose a continuous, infinite, periodic function with
a non-modulated RR\,Lyrae shape as a ``carrier wave''. 
This function is described by the frequency
$f_0$ and its harmonics, that is $c^*(t):=m(t)$ 
if $F_i=j f_0$ in Eq.~(\ref{old}). 

Although, the exact analytical Fourier spectra of
any of the modulated signals discussed in this paper 
can be calculated without any problems, at least in theory
(see also Appendix \ref{AAF}), to illustrate the different formulae 
synthetic light curves and their Fourier spectra
 were also generated and plotted.
An artificial light curve was constructed as a carrier wave with 
typical RR\,Lyrae parameters ($f_0=2$~d$^{-1}$ and its 9 harmonics)
on a 100 day long time span sampled by 5 min (insert in Fig.~\ref{carrier}).
The Fourier transform of such a signal is well-known (Fig.~\ref{carrier}):  
it consists of the transformed sinusoidal components given in 
Eq.~\ref{Fr_sinus}. (More precisely, due to the finite length 
of the data set and its sampling, the Fourier
transformations should always be multiplied by the Fourier transform
of the appropriate window function.)

Substituting $c^{*}(t)$ carrier wave 
into the definition of AM in Eq.~(\ref{mod_d}):
\begin{multline}\label{mod_star}
  m_{\mathrm{AM}}^*(t)=\left[ 1 + 
\frac{U_{\mathrm m}^* \left( t \right)}{U_{\mathrm c}^*} 
\right] c^*(t)={} \\
\left[ 1 + m_{\mathrm m}^*(t) \right]
\left[a_0 + \sum_{j=1}^n{a_j \sin(2\pi j f_0 t + \varphi_j)} \right].
\end{multline} 
Expression (\ref{mod_star}) 
describes a general amplitude modulated RR\,Lyrae
light curve, $U_{\mathrm m}^*(t)$ is the modulation signal, 
$U_{\mathrm c}^*$ is the amplitude of the non-modulated light curve. 
On the one hand, the non-zero constant term of $a_0$ 
is obligatory from mathematical point of view, otherwise the Fourier 
sum does not compose a complete set of functions.
On the other hand, this value represents the difference 
between the magnitude and intensity means.
More precisely, either we use physical quantities (viz. 
positive definite fluxes) or we transform normalized 
fluxes into magnitude scale. In this latter case the average of 
the transformed light curve differs from zero.
In this paper, for traditional purposes  we use the second approach.
For RR\,Lyrae stars the typical value of this difference is
about some hundredths of a magnitude ($a_0 \ll 1$). It is evident that 
this constant differs from the zero point of the light curve $A_{0}$
given in the apparent magnitude scale.

\begin{figure} 
\includegraphics[width=9cm]{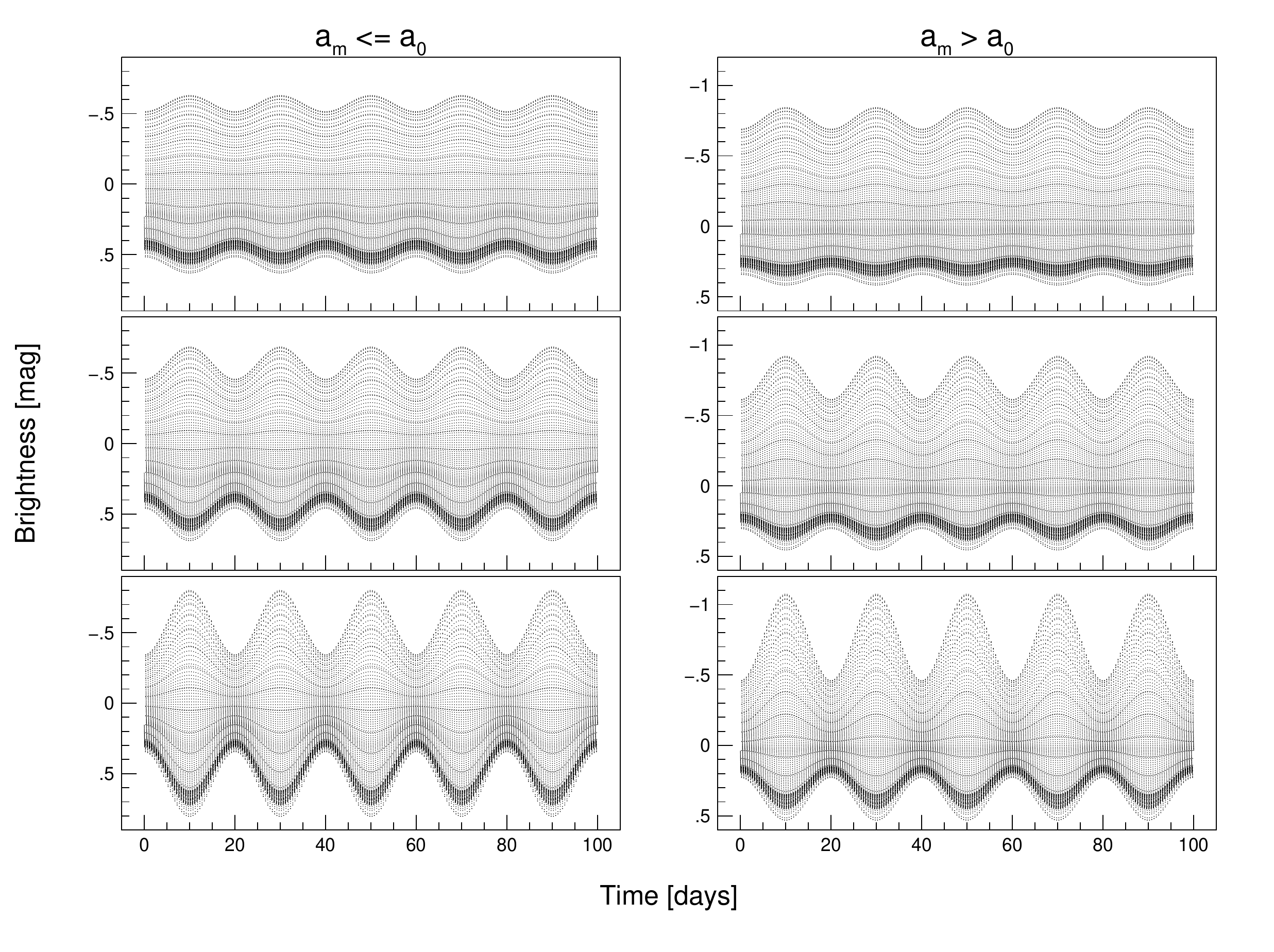}
\caption{Artificial light curves with a sinusoidal AM 
computed with the formula Eq.~(\ref{mod_star}). Left panels show
symmetrical modulation ($a_{\mathrm m} \le a_0$; $a_0=0.2$),
the right ones are asymmetrical ($a_{\mathrm m} > a_0$; $a_0=0.005$).
The modulation depth $h$ is increasing from the top 
to bottom as $h=0.1, 0.2, 0.4$; $f_{\mathrm m}=0.05$~d$^{-1}$ and 
$\varphi_{\mathrm m}=270$ deg are fixed.
}\label{am_sin}
\end{figure}

\subsubsection{Sinusoidal amplitude modulation}

 In the simplest case the modulation is sinusoidal:  
\begin{equation}\label{mod_sin}
 U_{\mathrm m}^*(t)=a_{\mathrm m}\sin (2\pi f_{\mathrm m} t + \varphi_{\mathrm m}).
\end{equation} 
Sample light curves obtained with this assumption from 
Eq.~(\ref{mod_star}) are shown in Fig.~\ref{am_sin}.
Introducing the modulation depth as $h=a_{\mathrm m} / U^*_{\mathrm c}$
the parameters were chosen as 
$a_0 \le U_{\mathrm c}^*$ and $a_{\mathrm m} \le a_0$, 
resulting in modulations symmetrical to an averaged value viz. a
horizontal line (left panels). 
Right panels show cases with higher modulation depths  
($a_{\mathrm m} > a_0$), where this symmetry is broken.
A common feature of these light curves is that
the maxima and minima of the envelope curves coincide in time. 
Furthermore, the average brightness of all light curves
vary with $f_{\mathrm m}$. It can be seen directly 
from Eq.~(\ref{mod_star}): 
the $m_{\mathrm m}^*(t)a_0$ term is responsible for this behaviour. 
That is, the found mean brightness ($\overline{V}$) 
variations during the Blazhko
cycle \citep{JurRR} is a natural consequence of the AM. 

\begin{figure} 
\includegraphics[width=9cm]{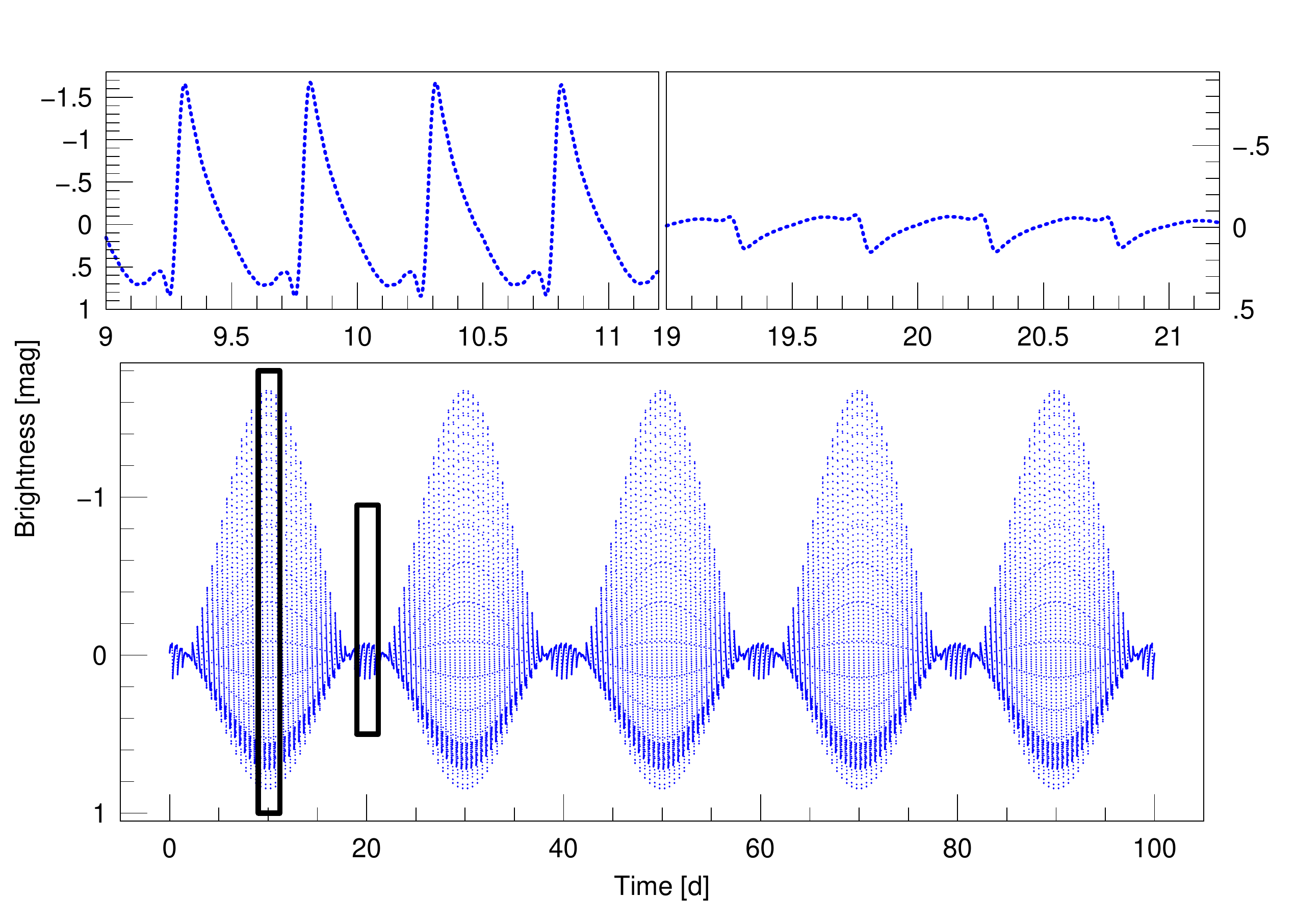}
\caption{Bottom panel: Artificial light curve with a sinusoidal AM 
computed with the formula (\ref{mod_star}). 
The modulation depth is $h=1.2$. Other parameters are
$a_0=0.01$, $f_{\mathrm m}=0.05$~d$^{-1}$ and
$\varphi_{\mathrm m}=270$ deg.
Top panels: Two-day zooms 
around a maximum (top left) and a minimum (top right) of the modulation
cycle.
}\label{inverse}
\end{figure}

There is a fascinating case,
 when the modulation is very strong i.e. when the 
modulation depth is $h > 1$. Beside the strong light curve changes 
 (Fig.~\ref{inverse}) in some Blazhko phases the shape 
of the light curve looks very unfamiliar 
(see top right panel in Fig.~\ref{inverse}). 
The relevance of this mathematical case is corroborated by the 
{\it Kepler\/} observation of V445\,Lyr that 
shows similar characteristics (fig~2. in \citealt{Ben10}). 

\begin{figure} 
\includegraphics[width=9cm]{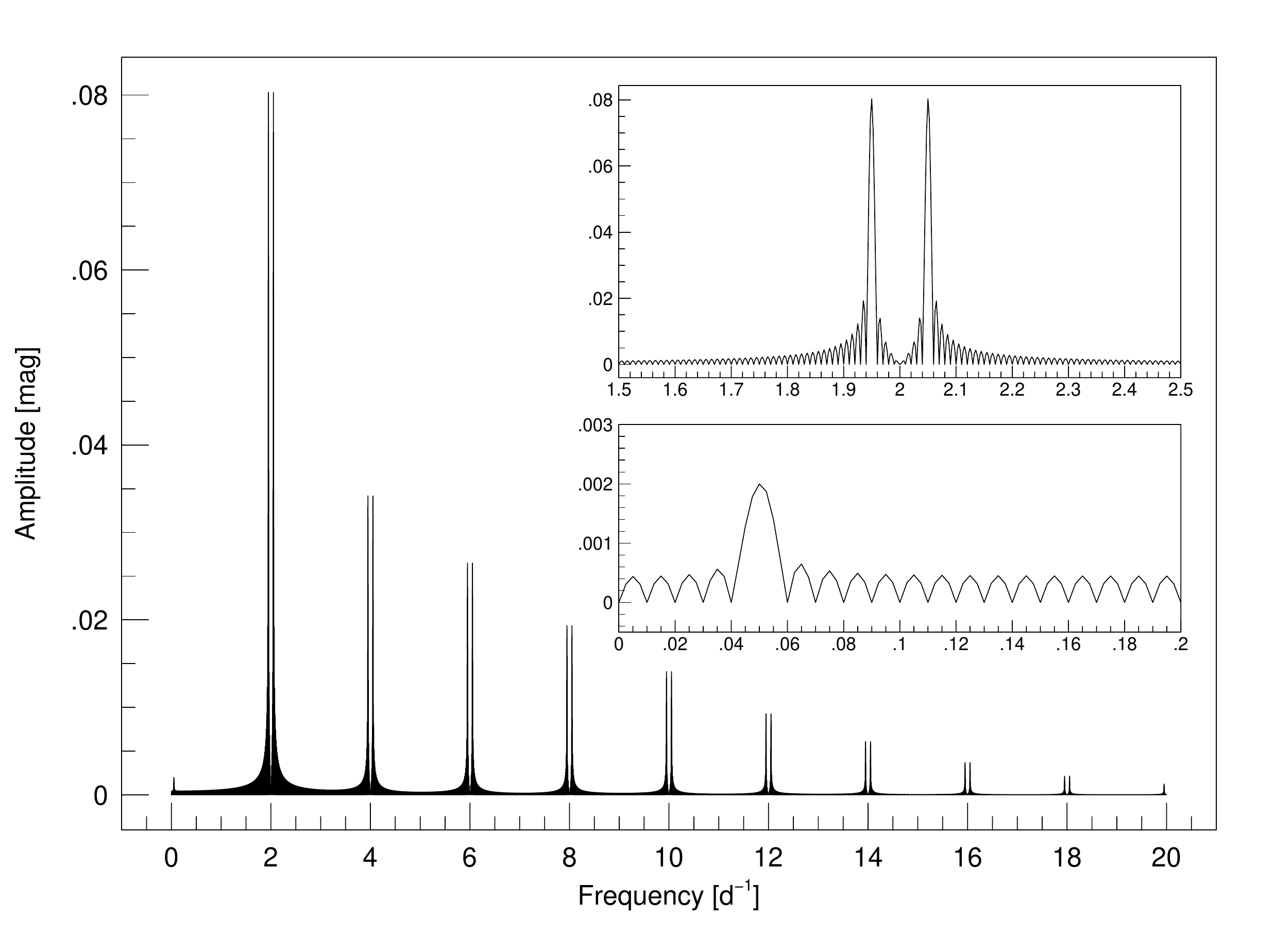}
\caption{
Fourier amplitude spectrum of the artificial sinusoidal AM
light curve in bottom right panel of Fig.~\ref{am_sin} 
after the data are prewhitened with the
main frequency and its harmonics.  Inserts are zooms around the 
positions of the main 
frequency $f_0=2$~d$^{-1}$ (top), and the modulation 
frequency $f_{\mathrm m}=0.05$~d$^{-1}$ (bottom),
respectively.
}\label{am_sinF}
\end{figure}

Using some trigonometrical relations, 
Eq.~(\ref{mod_star}) with (\ref{mod_sin}) 
can be converted to a handy sinusoidal decomposition form from where the 
Fourier spectrum is easily seen:  
\begin{multline}\label{mod_am_F}
m^*_{\mathrm{AM}}(t)=a_0 + h a_0 \sin \left( 2\pi f_{\mathrm m} t + 
\varphi_{\mathrm m} \right) + {} \\ 
\sum_{j=1}^{n} {a_j \sin \left( 2\pi j f_0 t+\varphi_j \right) } + {} \\ 
\frac{h}{2} \sum_{j=1}^{n} a_j \left\{ 
\sin \left[ 2\pi \left( j f_0 - f_{\mathrm m} \right) t 
+\varphi^{-}_j \right] + \right. \\
\left. \sin \left[ 2\pi \left( j f_0 + f_{\mathrm m} \right) t 
+\varphi^{+}_j \right] \right\},
\end{multline}
where $\varphi_j^{-}=\varphi_j-\varphi_{\mathrm m}+\pi /2$, 
 $\varphi_j^{+}=\varphi_j+\varphi_{\mathrm m}-\pi /2$. 
The Fourier spectrum of such an AM signal  
is familiar for RR\,Lyrae stars' experts 
(see also Fig.~\ref{am_sinF}).
It consists of the spectrum of the non-modulated star (third term) as 
in Fig.~\ref{carrier} and two equidistant side peaks 
around each harmonic (last term). The amplitudes of the pairs of 
side peaks are always equal: $A(j f_0 \pm f_{\mathrm m}) \sim  a_j h /2$. 
The second term  in Eq.~(\ref{mod_am_F}), 
that causes the average brightness variation, produces
the frequency $f_{\mathrm m}$ itself in the spectrum. 
The Blazhko modulation frequency
is always found in observed data sets extended enough 
(see e.g. \citealt{Kov95, Nag98, JurRR, JurMW, Cha10, Por10, Kol11}).  

It is a long-standing question whether there is any Blazhko phase
where the modulated light curve is identical to the unmodulated one
(see \citealt{Jur02} and further references therein).
In this simplest case the answer is easy. It happens if
the second and fourth terms in (\ref{mod_am_F}) disappear 
simultaneously, namely, in the zero points of 
the modulated sinusoidal function at 
$t= (k\pi -\varphi_{\mathrm m})/ (2\pi f_{\mathrm m})$, $k$ is 
an arbitrary integer.

The number of used parameters 
for solving such a light curve (Fig.~\ref{am_sin}) in 
the traditional way (according to Eq.~\ref{old}) 
is $6n+5$, where $n$ denote the number of detected 
harmonics including the main frequency. 
The necessary number of parameters in our handling is $2n+5$.
The modulation is described by 3 parameters 
($f_{\mathrm m}$, $a_{\mathrm m}$, $\varphi_{\mathrm m}$)
 as opposed to the traditional framework where this 
number is $4n+3$.  

\subsubsection{Non-sinusoidal AM} 

\begin{figure} 
\includegraphics[width=9cm]{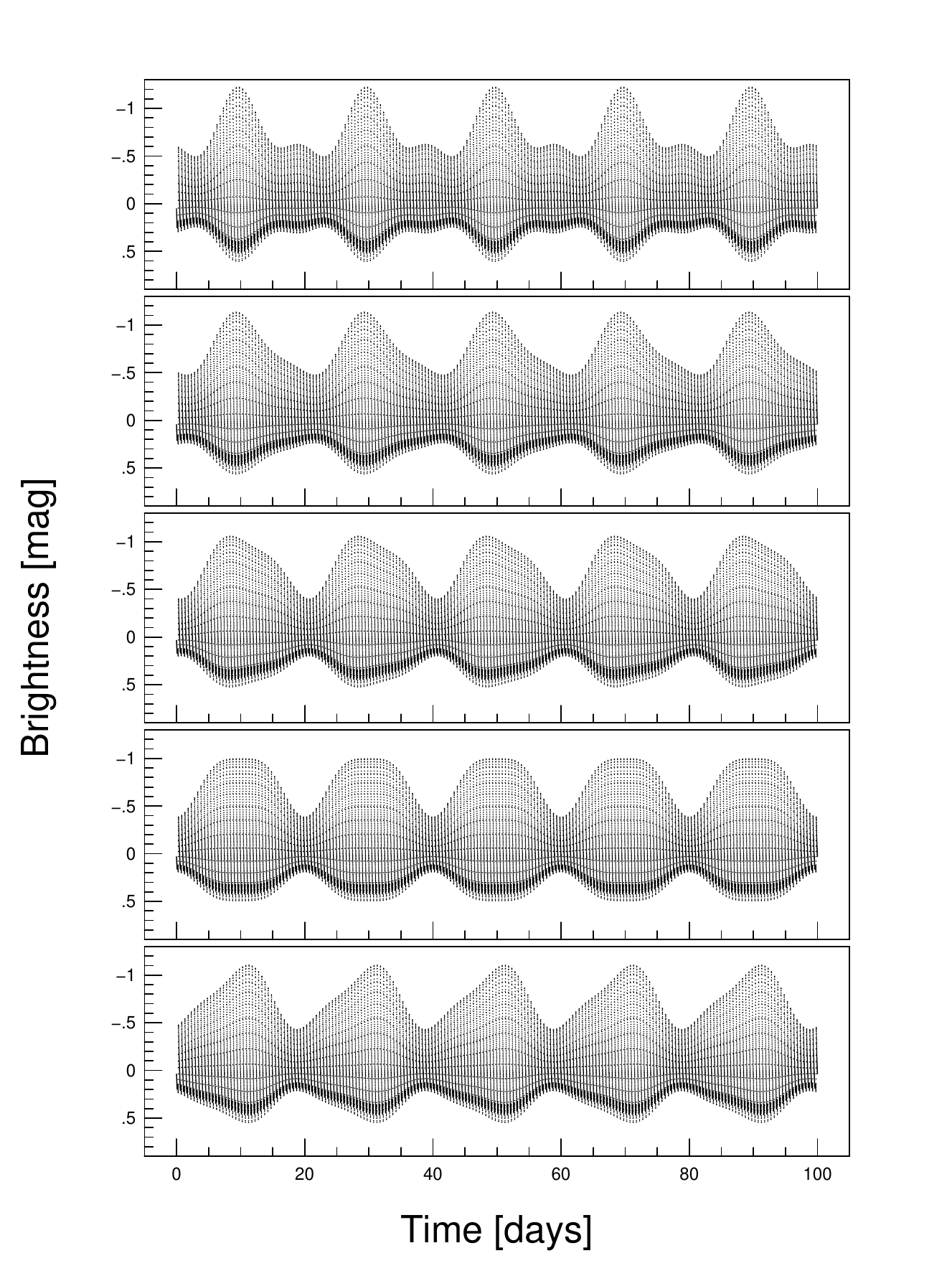}
\caption{Synthetic light curves of non-sinusoidal AM 
signals computed by the formula (\ref{mod_AM}).
A two-term sum of modulation signal was assumed: $a^{\mathrm A}_1=0.01$, 
 $a^{\mathrm A}_2=0.2$~mag;  $\varphi^{\mathrm A}_1=270$~deg 
are fixed and the
phase of the second modulation term varies from top to bottom as:
$\varphi^{\mathrm A}_2=110$, 140, 220, 270, 360, respectively.   
}\label{am_nsinlc}
\end{figure}

As a next step, we assume the
modulation function $m_{\mathrm m}^*(t)$ to be an arbitrary periodic
signal represented by a Fourier sum with a constant frequency 
$f_{\mathrm m}$. Substituting it into Eq.~(\ref{mod_star}) we get 
\begin{equation}\label{mod_AM}
m_{\mathrm{AM}}^*(t)=\left[ a^{\mathrm A}_0+ 
\sum_{p=1}^q{a^{\mathrm A}_p 
\sin\left( 2\pi p f_{\mathrm m} t + \varphi^{\mathrm A}_p \right)} 
\right] c^*(t),
\end{equation}
where constants are defined by 
$a^{\mathrm A}_0=1+(a_0^m/ U_{\mathrm c}^*)$, 
and $a^{\mathrm A}_p=a_p^m/U_{\mathrm c}^*$.
From now on, the upper index A denotes the AM parameters.  
Some typical light curves are shown is Fig.~\ref{am_nsinlc}.
It is evident that their envelope curves are non-sinusoidal 
and their shapes depend
on the actual values of $a^{\mathrm A}_p$ and $\varphi^{\mathrm A}_p$.
The maxima and minima of these envelope curves occur, however,
at the same Blazhko phase as in the previous sinusoidal cases. 
Rewriting of (\ref{mod_AM}) similarly to Eq.~(\ref{mod_am_F})
but in a more compact form yields 
\begin{multline}\label{mod_am_nsinF}
m^*_{\mathrm{AM}}(t)=  
\sum_{p=0}^{q}
\sum_{j=0}^{n} \frac{a^{\mathrm A}_p}{2} a_j  
\sin \left[ 2\pi \left( j f_0 \pm p f_{\mathrm m} \right) t 
+\varphi^{\pm}_{jp} \right],
\end{multline}
where the two sinusoidal terms appearing analogously to  
Eq.~(\ref{mod_am_F}) are formally unified into one formula and denoted 
by $\pm$ signs;
$\varphi^{+}_{jp}=\varphi_j+\varphi^{\mathrm A}_p-\pi /2$;
 $\varphi^{-}_{jp}=\varphi_j-\varphi^{\mathrm A}_p+\pi /2$. 
The arbitrary constants
are chosen to be $\varphi^{\mathrm A}_0:=\varphi_{0}:=\pi /2$.

\begin{figure} 
\includegraphics[width=9cm]{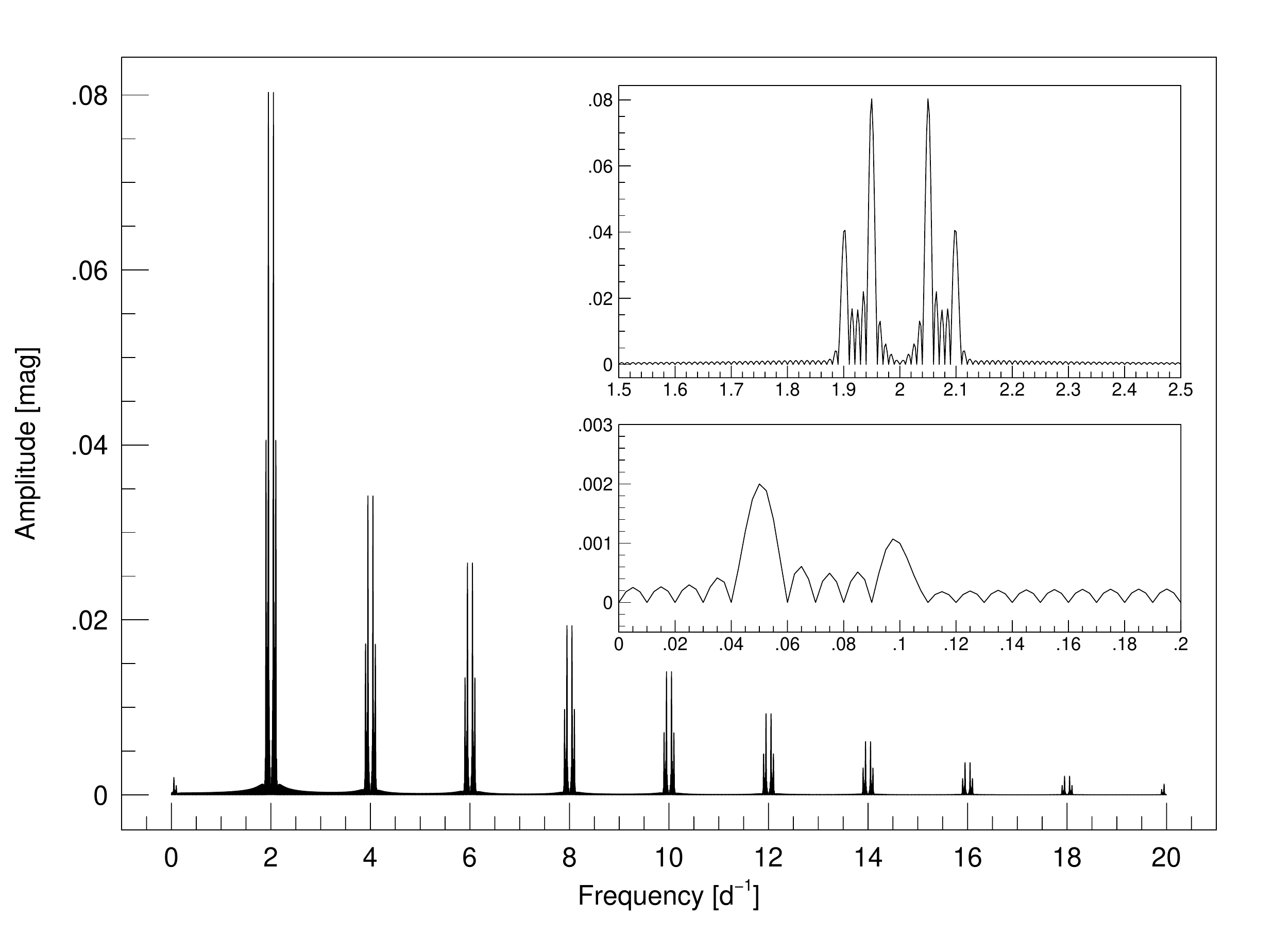}
\caption{
Fourier amplitude spectrum of the synthetic non-sinusoidal AM
light curves in Fig.~\ref{am_nsinlc} after the data 
were prewhitened with the
main frequency and its harmonics.  Inserts are zooms around the positions of the main 
frequency $f_0=2$~d$^{-1}$ (top), and 
the modulation frequency $f_{\mathrm m}=0.05$~d$^{-1}$ (bottom),
respectively.
}\label{am_nsinF}
\end{figure}

By investigating the Fourier amplitude of the side peaks we found that 
$A(jf_0\pm pf_{\mathrm m})/A(jf_0) \sim a^{\mathrm A}_p$.
(i) The amplitude ratio of the side peaks of
a given order and the central peak is constant.
(ii) The commonly used amplitude ratio
$A(jf_0\pm pf_{\mathrm m})/A(f_0\pm pf_{\mathrm m}) \sim a_j /a_1$
vs. harmonic order is the same as the amplitude ratio of the main 
frequency $A(jf_0)/A(f_0) \sim a_j /a_1$ vs. harmonic order. 
(iii) Since the same coefficient
$a^{\mathrm A}_p$ belongs to both side peaks (at $\pm p$),
the amplitudes of left-hand-side and right-hand-side peaks 
are the same. According to this, the generated Fourier spectrum 
(Fig~\ref{am_nsinF}) now shows symmetrical multiplet structure 
of peaks around the  main frequency and its harmonics 
($ j f_0 \pm p f_{\mathrm m}$). 
Each multiplet structure is the same at each harmonic 
order, that is  the number of the side peaks, their
frequency differences and amplitude ratios 
to their central peaks are the same.
It is important to note that the number of side peaks 
(in one side) is equal to $p$.  
In addition, the harmonic components of the modulation frequency 
$p f_{\mathrm m}$ also appear. (This can be obtained
from Eq.~(\ref{mod_am_nsinF}) if $j=0$.)

Such a phenomenon was undetected in the observed data of Blazhko
stars until recently. \cite{Hur08} found equidistant quintuplets 
in the spectrum of RV\,UMa for the first time. Besides triplets and quintuplets,
sextuplet structures were also found by \cite{JurMW}
in the spectrum of MW\,Lyr, while \cite{Cha10} detected even 8th order
(sepdecaplet) multiplet frequencies in the spectrum of {\it CoRoT\/} data of V1127\,Aql.
According to Sec.~\ref{Basic_FM}, the angle modulations cause 
infinite numbers of side peaks around each harmonic, therefore, 
the origin of the observed multiplets 
as a non-sinusoidal amplitude modulation is certain for 
those cited cases (e.g. MW\,Lyr and V1127\,Aql), where 
the harmonics of the modulation frequency are also detected. 

In searching for a Blazhko phase where 
the modulated and carrier waves are identical we concluded
that the modulation terms can only be entirely disappearing from the
formula (\ref{mod_am_nsinF}) if $a^{\mathrm A}_0=1$ ($a^{\mathrm m}_0=0$) 
is true, otherwise no such Blazhko phase exists. 
This necessary condition 
is complemented by an additional one: the modulation 
virtually disappears in the moments when 
$\sum_{p=1}^{q}a_ja_p\sin [2\pi (j f_0 \pm p f_{\mathrm m})t + 
\varphi^{\mp}_{jp}]=0$. The sum has either zero or infinite numbers of
zero points depending on the values of the parameters 
$a^{\mathrm A}_p$, $\varphi^{\mathrm A}_p$.
That is, generally there are no such phases where a 
non-sinusoidal AM light curve and its carrier wave are identical. 

The necessary number of the parameters for a light curve 
fit of (\ref{old}) and (\ref{mod_AM}) 
is $(2q+1)2n+2q+3$ and $2n+2q+3$, respectively. 
Here $n$ denotes the total number
of used harmonics including the main frequency and $q$ is the order of 
side peak structures as above, 
(i.e. $q=1$ means triplets, $q=2$ quintuplets, etc). In the 
traditional description each additional side peak order 
increased the number of 
terms by $4n+2$ as opposed to our method, where this increment is only 2.

\subsubsection{Parallel AM modulation}

Multiperiodic modulation was suspected in XZ\,Cyg \citep{Lac04},
UZ\,UMa \citep{Sod06}, SU\,Col \citep{Szc07}, and LS\,Her \citep{Wil08}.
The Blazhko RR Lyrae stars of the MACHO and OGLE surveys \citep{Alc00, MP03}
that have unequally spaced triplet structures in their Fourier spectra
are possibly also multiperiodically modulated variables. 
CZ\,Lac \citep{Sod10} 
is the first Blazhko star with multiperiodic modulation 
where both modulation periods are identified. 
Not only modulation side peaks but linear combinations
of the modulation frequencies also appear. 
Signs of multiple modulation were discovered
in {\it Kepler\/} data of V445\,Lyr \citep{Ben10}.
There are numerous possibilities for creating a 
multiply modulated light curve. Let us review some of them. 

\begin{figure} 
\includegraphics[width=9cm]{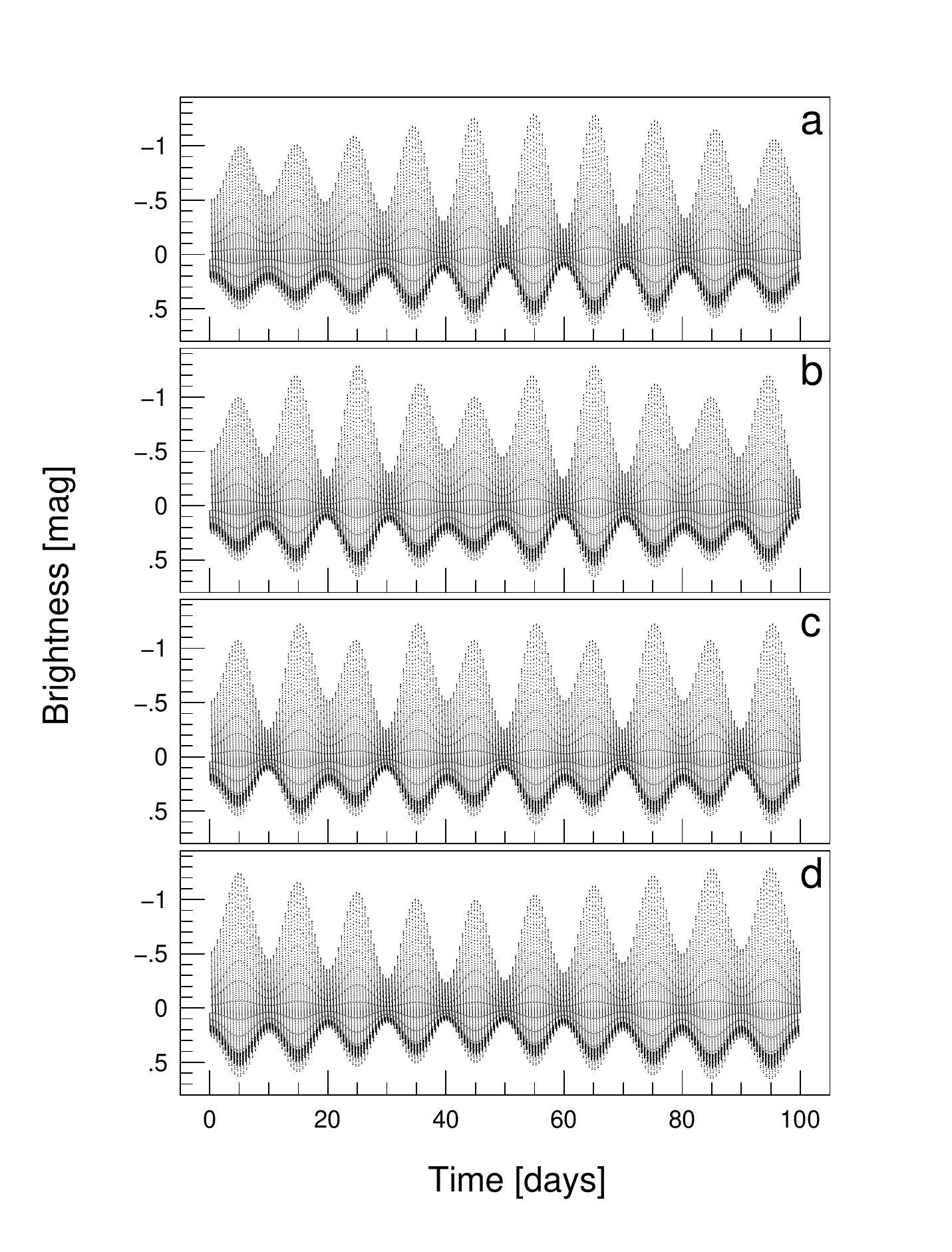}
\caption{Artificial light curves calculated with two 
independent sinusoidal 
AM modulations according to the formula of (\ref{mod_MAM}).
The fixed parameters were $a_0=0.01$, $\hat{a}_{11}=0.5$,
$\hat{a}_{12}=0.2$~mag, $\hat{f}^1_{\mathrm m}=0.1$~d$^{-1}$,
$\hat{\varphi}_{11}=270$, $\hat{\varphi}_{12}=120$~deg, where 
$\hat{f}^2_{\mathrm m}$ changes from top to bottom as
0.09, 0.075, 0.05 and 0.01~d$^{-1}$, respectively.
}\label{am_pd}
\end{figure}

The most simple case is a  natural generalization 
of Eq.~(\ref{mod_AM})  when the modulation signal is 
assumed to be a sum of signals with different 
$\hat{f}^{r}_{\mathrm m}$, where 
$r=1, 2, \dots$ signs constant frequencies.
Let signals be independent, i.e., the modulation signal
consists of linearly superimposed waves.
In this case, Eq.~(\ref{mod_AM}) reads as:
\begin{equation}\label{mod_MAM}
m_{\mathrm{AM}}^*(t)=\left[ \hat{a}_0 + 
\sum_{r=1}^s{
\sum_{p=1}^{q_r}{\hat{a}_{pr} 
\sin\left( 2\pi p
\hat{f}^r_{\mathrm m} t + \hat{\varphi}_{pr} \right)}} 
\right] c^*(t),
\end{equation}
where 
$\hat{a}_0=1+\sum_{r=1}^s a^m_{0r} / {U^*_{\mathrm c}}$,
and $\hat{a}_{pr}=a^m_{pr}/ U^*_{\mathrm c}$.
This formula is demonstrated in Fig.~\ref{am_pd}.
In this figure, only two modulation waves 
are taken into account and the only varied parameter is the
frequency of the second modulation is $\hat{f}^2_{\mathrm m}$.
When the two modulation frequencies are comparable
($\hat{f}^1_{\mathrm m}=0.1$~d$^{-1}$ and 
$\hat{f}^2_{\mathrm m}=0.09$~d$^{-1}$ 
in panel a the envelope shape of the light curve 
shows the well-known beating phenomenon. Here the 
beating period is 200 days, albeit the modulation periods
are close to the shortest known ones. It is easy to
understand that the observations taken on a moderate time span
often detect only the gradual increase or decrease of
 the amplitude of the Blazhko cycles. In panel b of Fig.~\ref{am_pd} 
$\hat{f}^2_{\mathrm m}$ was set to 0.075~d$^{-1}$, that is
the ratio of the modulation frequencies is 4:3, similarly to
the case of CZ\,Lac during its second observed season in \cite{Sod10}.
The amplitude changes of the consecutive Blazhko cycles 
need well-covered long-term time series observations, otherwise the 
interpretation becomes difficult. Panel c in
Fig.~\ref{am_pd} shows the case where the frequency of the 
second modulation is half of the first one 
($\hat{f}^2_{\mathrm m}=0.05$~d$^{-1}$). These specially selected
values cause alternating higher and lower Blazhko cycles.
The exact 2:1 ratio between the two modulation frequencies 
leads to the same result as a two-term non-sinusoidal modulation 
in Eq.~(\ref{mod_AM}) (see also top panel in Fig.~\ref{am_nsinlc}).
The bottom panel in Fig.~\ref{am_pd} shows a
case where the second modulation has a much longer period than the
primary Blazhko cycle. In a first inspection the top and bottom 
panels are very similar apart from a phase shift. 

To reveal the real situation we need to compare their Fourier spectra
starting with
\begin{equation}\label{mod_MAMF}
m_{\mathrm{AM}}^*(t)=  
\sum_{r=1}^s{ \sum_{p=0}^{q_r}{\sum_{j=0}^{n}{
\frac{\hat{a}_{pr}}{2} a_j \sin\left[ 2\pi \left( j f_0 \pm
p \hat{f}^r_{\mathrm m} \right) t 
+ \hat{\varphi}^{\pm}_{jpr} \right] }}}.
\end{equation}
Where the constants are chosen similarly to 
(\ref{mod_am_nsinF}): 
$\hat{\varphi}^{-}_{jpr}=\varphi_j-\hat{\varphi}_{pr}+\pi /2$;
$\hat{\varphi}^{+}_{jpr}=\varphi_j+\hat{\varphi}_{pr}-\pi /2$;  
$\varphi_0=\hat{\varphi}_{0r}=\pi /2$.
It is easy to see that 
the Fourier spectrum of expression (\ref{mod_MAMF}) 
contains $s$ sets of side peaks shown in Fig.~\ref{am_nsinF}.
The qualitative structure of these sets is the same. It consists of
the carrier's spectrum ($j f_0$), the peaks of
the different modulation frequencies and their harmonics 
($p \hat{f}^r_{\mathrm m}$),
and the side peaks around the main frequency and its harmonics:
$j f_0\pm p \hat{f}^r_{\mathrm m}$, 
where $p=1,2$, \dots, $q_r$, and $r=1,2,\dots, s$.
Due to the independence of the modulation 
waves no further side peaks appear.

\subsubsection{Modulated modulation -- the AM cascade}

\begin{figure} 
\includegraphics[width=9cm]{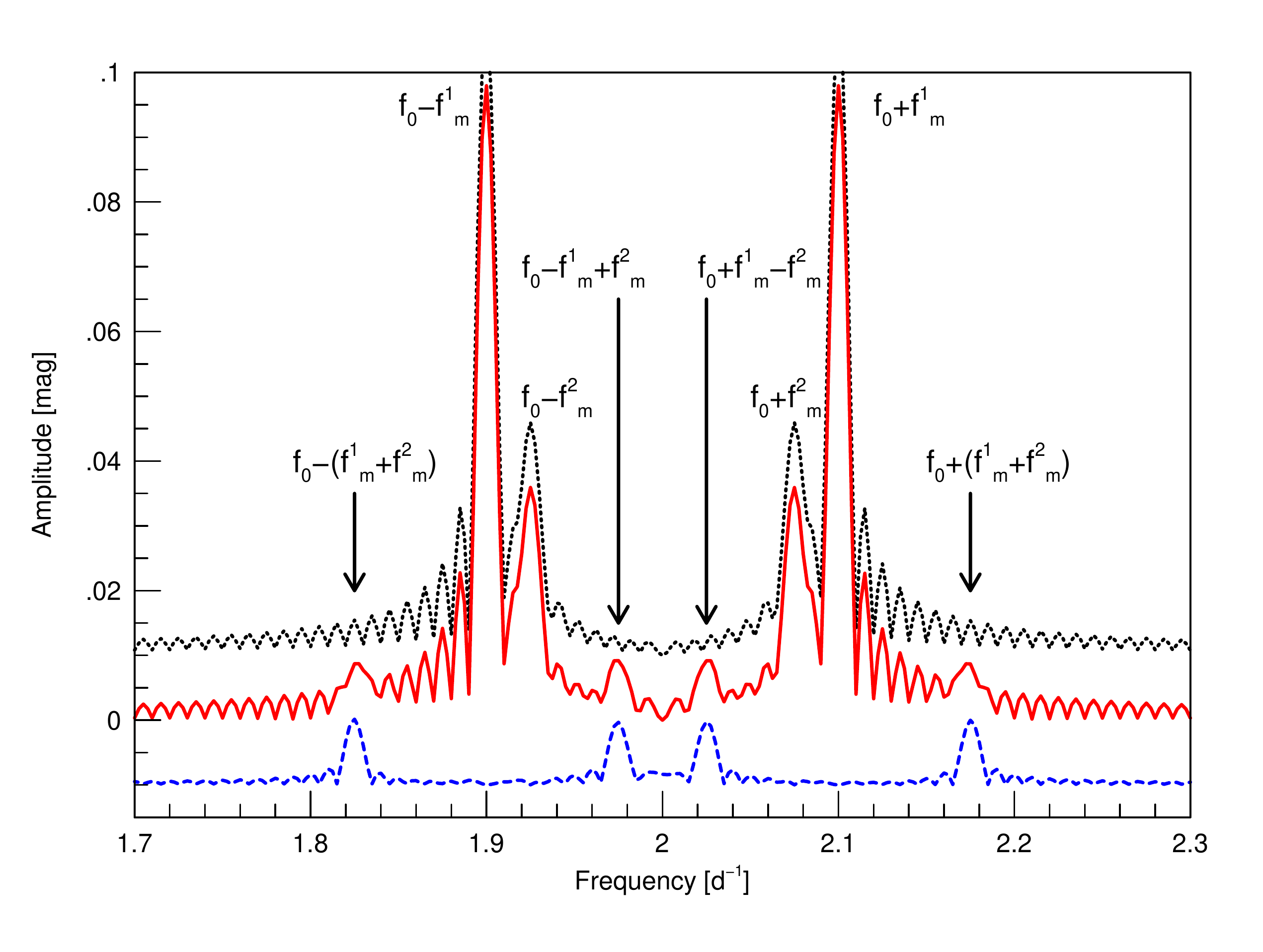}
\caption{Fourier amplitude spectra of the artificial
cascade AM light curves with parameters of panel b 
in Fig.~\ref{am_pd} -- (black) dotted line --
and its cascade equivalent -- (red) continuous line. 
The spectra show the interval around the main pulsation frequency
after the data are prewhitened with it.
The (blue) dashed line shows the spectrum of cascade case after 
the side peaks $f_0\pm f^1_{\mathrm m}$, 
$f_0\pm f^2_{\mathrm m}$ are also removed (for better visibility
the top and bottom spectra are shifted with +0.01 and $-$0.01~mag, 
respectively).
}\label{am_casf}
\end{figure}

It is hard to imagine, however, that in a real stars's case,  
the different modulating waves are independently superimposed
without any interactions. Let us investigate 
the possibility of the modulated modulation: the cascade.
In other words, the modulation signal is composed of recursively
modulated waves as 
\begin{multline}
c^{(1)}(t):=c^*(t), \ \ m^{(1)}_{\mathrm m}(t)=m^*_{\mathrm m}(t), \\
\shoveleft{c^{(2)}(t):=m_{\mathrm{AM}}^{(1)}(t)=[1+m^{(1)}_{\mathrm m}(t)]c^{(1)}(t),} \\ 
m_{\mathrm{AM}}^{(2)}(t)=[1+m^{(2)}_{\mathrm m}(t)]c^{(2)}(t), \dots, \\
m_{\mathrm{AM}}^{(s)}(t)=[1+m^{(s)}_{\mathrm m}(t)]c^{(s)}(t).
\end{multline}
\begin{equation}\label{mod_MAM2}
m_{\mathrm{AM}}^*(t)=\prod_{r=1}^s{\left[ 
\tilde{a}_{0r}+ 
\sum_{p=1}^{q_r}{\tilde{a}_{pr} 
\sin\left( 2\pi p \tilde{f}^r_{\mathrm m} t + \tilde{\varphi}_{pr} 
\right)} 
\right]} c^*(t),
\end{equation}
where $\tilde{a}_{0r}=1+a^m_{0r}/U^*_{cr}$, $\tilde{a}_{pr}=a^m_{pr}/U^*_{cr}$.
$U^*_{cr}$ denotes the amplitude of the $r$th carrier wave $c^{(r)}(t)$.
On the basis of a  visual inspection,
there are imperceptible differences among the light curves produced 
by this expression (\ref{mod_MAM2}) 
and those that can be seen in Fig.~\ref{am_pd}. 
The Fourier spectrum, however, contains additional peaks 
at the linear combinations of $f_0$ and $\tilde{f}^{r}_{\mathrm m}$
as it is shown in Fig~\ref{am_casf}. To understand this spectrum
we generate Eq.~(\ref{mod_MAM2}) in a form similar to Eq.~(\ref{mod_MAMF}).
\begin{multline}\label{cascade_amF}
m_{\mathrm{AM}}^*(t)=  
\sum_{r=1}^s \sum_{p=0}^{q_r} \sum_{j=0}^{n} \frac{1}{2}
\left(
\prod_{\stackrel{r'=1}{r'\ne r}}^{s} \tilde{a}_{0r'} \right) 
\tilde{a}_{pr} a_j \cdot {} \\
\shoveright{
\sin\left[ 2\pi \left( j f_0 \pm
p \tilde{f}^r_{\mathrm m} \right) t 
+ \tilde{\varphi}^{\pm}_{jpr} \right] +} {} \\
\sum_{k=2}^{s} \sum_{r\in \mathcal{R}_k}^{} \sum_{p=1}^{q_r}
a_0\prod_{r'\in {\mathcal{R}}^{\mathrm C}_k}^{}\tilde{a}_{0r'}
\left(\prod_{r}^{}\tilde{a}_{pr} \right)
\mathcal{S}_k(\boldsymbol{\alpha}) + \\
\sum_{k=2}^{s} \sum_{r\in \mathcal{R}_k}^{} \sum_{p=1}^{q_r}
\sum_{j=1}^{n}
a_j\prod_{r'\in {\mathcal{R}}^{\mathrm C}_k}^{}\tilde{a}_{0r'}
\left(\prod_{r}^{}\tilde{a}_{pr} \right)
\mathcal{S}_{k+1}(\boldsymbol{\beta}),
\end{multline}
$\tilde{\varphi}^{-}_{jpr}=\varphi_j-\tilde{\varphi}_{pr}+\pi /2$;
$\tilde{\varphi}^{+}_{jpr}=\varphi_j+\tilde{\varphi}_{pr}-\pi /2$;  
$\tilde{\varphi}_{0r}=\pi /2$.
An index set $\mathcal{R}_s$ is defined so that it contains all $r$
indices from $r=1,2,\dots,s$. Index sets $\mathcal{R}_k$ means 
all subsets of $\mathcal{R}_s$ what contains $k$ elements.
Therefore, the total number of $\mathcal{R}_k$ sets is $\binom{s}{k}$.
The sums over $r \in \mathcal{R}_k$ mean a sum over the all
possible combinations of the $k$ number of different indices $r$.
Similarly $r'$ always runs over the complement of a set 
$\mathcal{R}^{\mathrm C}_k$ of the
actual $\mathcal{R}_k$. The functions $\mathcal{S}_k$ build 
up sums of sinusoidal functions 
(see for the definition and further details in Appendix \ref{AP-S})
of the linear combinations of $k$ angles. Here the components of
the $\boldsymbol{\alpha}$ and $\boldsymbol{\beta}$ vectors are 
$\alpha_r=2\pi p \tilde{f}^{r}_{\mathrm m}t +
\tilde{\varphi}_{pr}$, $r \in \mathcal{R}_k$,
$\beta_{k+1} = 2\pi j f_0 t + \varphi_j$ and 
$\beta_r=\alpha_r$.

For better comparison we present the formula in the form
of Eq.~(\ref{cascade_amF}) instead of the most possible compact one. 
Although Eq.~(\ref{cascade_amF}) seems to be complicated, the 
meaning of each term is simple:
the first term can be directly compared to the linearly superimposed
case Eq.~(\ref{mod_MAMF}). It produces all the peaks in the Fourier 
spectrum appearing in that case: the main pulsation frequency 
and its harmonics $jf_0$, modulation frequencies and its harmonics
$p\tilde{f}^r_{\mathrm m}$, the side frequencies around the
main frequency and its harmonics ($jf_0\pm p\tilde{f}^r_{\mathrm m}$).
The second sum in Eq.~(\ref{cascade_amF}) 
belongs to the peaks of all the 
possible linear combinations of $p\tilde{f}^r_{\mathrm m}$,
whilst the last term is responsible for the many linear 
combination frequencies around the main frequency and its harmonics 
(Fig.~\ref{am_casf}). The latter two types of combination terms
were detected by \cite{Sod10} in the Fourier spectrum 
of CZ\,Lac, the only well-studied multiply modulated RR\,Lyrae star.

Long-term secondary changes in Blazhko cycles can be explained by
the variable strength of the modulation. To formulate this 
assumption we arrived at
\begin{equation}\label{mod_var}
m_{\mathrm{AM}}^*(t)=\left\{ 1+ \left[ 1+m'_{\mathrm m}(t) \right] m''_{\mathrm m}(t) \right\} c^*(t)
\end{equation}
The formula (\ref{mod_var}) can be considered as a special
case of (\ref{mod_MAM2}) when $s=2$ and $\tilde{a}_{01}=0$.

\subsection{Blazhko stars with FM}\label{Bl_FM}

We remind the reader of the possible absence of 
real Blazhko stars with pure AM that was mentioned in the 
introductory paragraphs of Sec.~\ref{Sec_AM}.
The only difference between pure AM and FM cases is that
RR\,Lyrae stars showing pure PM/FM are much more rarely 
reported than pure AM ones, but there are some examples 
(e.g. \citealt{Kur00, Der04}). 

How can the formalism  
discussed in Sec.~\ref{Basic_FM}  be applied to RR\,Lyrae stars?
Let us assume the same carrier wave as in the case of AM,
but here the instantaneous frequency $f(t)$ is denoted as
$f_0+m^*_{\mathrm m}(t)$, and $m^*_{\mathrm m}(t)$ is an arbitrary
(bounded) modulation signal. 
\begin{equation}\label{mod_fm}
  m_{\mathrm{FM}}^*(t)= a_{0} + 
\sum_{j=1}^n{a_j 
\sin \left\{ 2\pi j \left[ f_0+m^*_{\mathrm m}(t) \right] t 
+ \varphi_j \right\} }.
\end{equation}  
Expression (\ref{mod_fm})
describes a general frequency modulated RR\,Lyrae light curve.

\begin{figure}
\includegraphics[width=9cm]{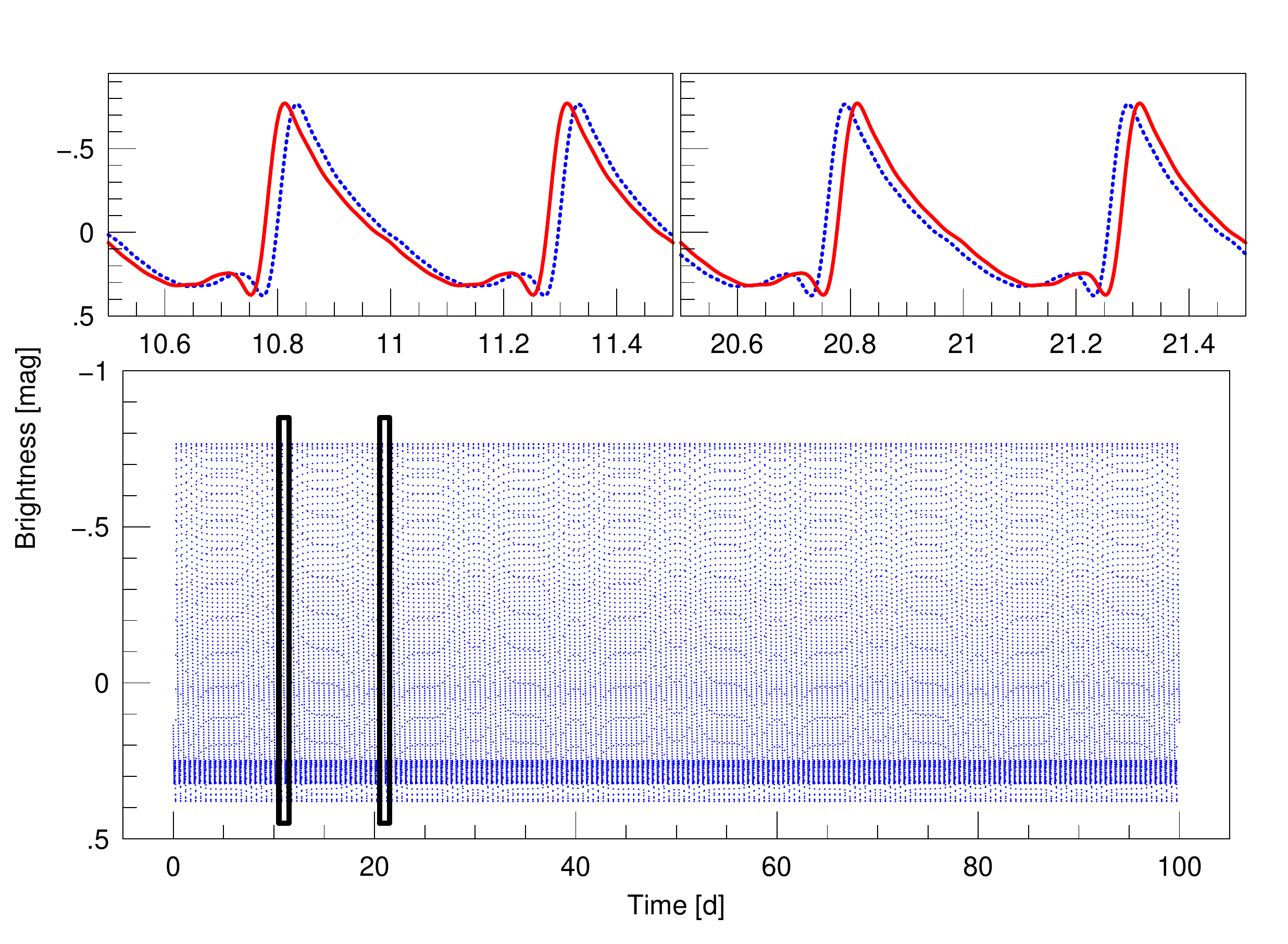}
\caption{
Bottom: Artificial FM light curve produced with 
sinusoidal modulation according to Eq.~(\ref{mod_fm_sin}). 
Fourier parameters of the ``carrier light curve'' are the same as
before, $a^{\mathrm F}=0.279; \varphi^{\mathrm F}=0$.
Boxes show the locations of the top panels.
Top panels: One-day zooms from two different phases of
the modulation cycle. 
The non-modulated ``carrier'' light curve is shown by a (red) continuous
lines while (blue) dotted lines denotes FM signal.
The periodic phase shift (viz. PM) caused by FM can be clearly identified.
}\label{Fig_fm_sin}
\end{figure}
\begin{figure} 
\includegraphics[width=9cm]{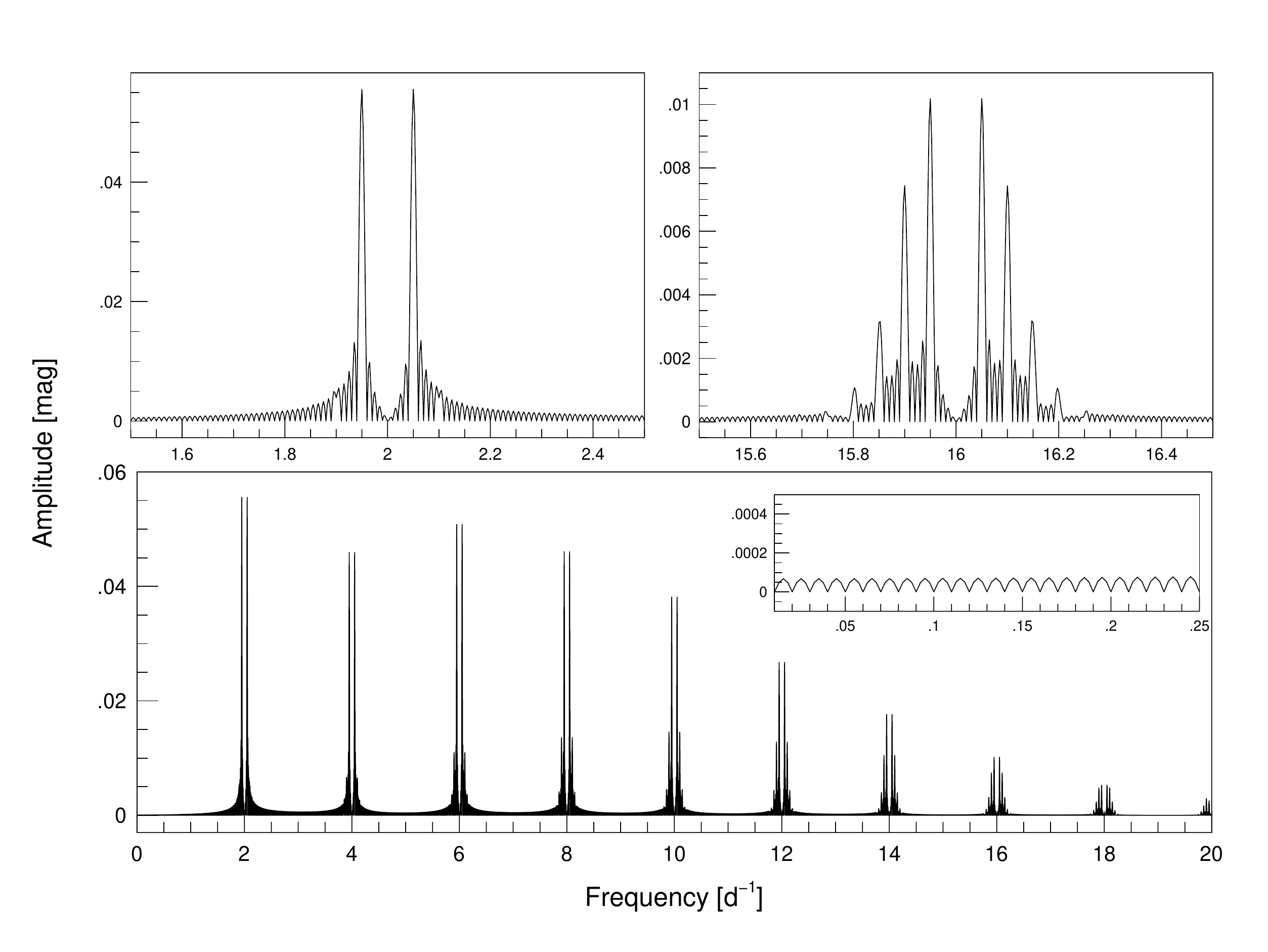}
\caption{Bottom:
Fourier amplitude spectrum of the artificial sinusoidal FM
light curve in Fig.~\ref{Fig_fm_sin} after the data are prewhitened 
with the main frequency and its harmonics.  
Top panels are zooms around the positions of the main 
frequency $f_0=2$~d$^{-1}$ (top left), and its 7th harmonics 
$8f_0=16$~d$^{-1}$ (top right), respectively.
The modulation frequency $f_m$ is missing from the spectrum (insert).
}\label{fm_sinF}
\end{figure}

\subsubsection{The sinusoidal FM}\label{Sec_sinfm}

When the modulating function is sinusoidal and 
expressed in the same form as (\ref{mod_sin}), 
Eq.~(\ref{mod_fm}) becomes   
\begin{multline}\label{mod_fm_sin}
  m_{\mathrm {FM}}^*(t)= a_{0} + {}\\
\sum_{j=1}^n{a_j 
\sin \left[ 2\pi j f_0 t + j a^{\mathrm F} 
\sin \left( 2\pi f_{\mathrm m} t + \varphi^{\mathrm F} \right) 
+ \varphi_j \right]},
\end{multline}  
where $a^{\mathrm F}=a_{\mathrm m} /  f_{\mathrm m}$, 
$\varphi^{\mathrm F}=\varphi_{\mathrm m}+\pi /2$
and the upper index F marks the parameters of FM. 
The amplitude of this signal is determined by the Fourier
amplitudes $a_j$ of the carrier signal, hence no amplitude
changes are present. In the bottom panel of Fig.~\ref{Fig_fm_sin}
a simulated light curve is shown. It is evident that there is 
no amplitude change.
Two-day zooms from two different phases of
the modulation cycle are shown in the top panels.
The periodic phase shift caused by FM can be identified well:
in the left-hand-side panel the non-modulated light curve is
to the left from the FM light curve whilst in the right-hand side
panel the situation is opposite.  

Using Chowning's relation (\ref{mod_Chowning}) 
we get from (\ref{mod_fm_sin}):
\begin{multline}\label{mod_fm_sinF}
  m_{\mathrm {FM}}^*(t)= a_{0} + {} \\
\sum_{j=1}^n 
{\sum_{k=-\infty}^{\infty}
{a_j J_k\left( j a^{\mathrm F} \right) 
\sin \left[ 2\pi \left( j f_0 + k f_{\mathrm m} \right) t  
+ k \varphi^{\mathrm F} + \varphi_j \right] }}.
\end{multline}  
This equation shows the main characteristics of the Fourier spectrum
(Fig.~\ref{fm_sinF}).
It consists of peaks at $f_0$ and at its harmonics $j f_0$ and 
each of them is surrounded by side peaks at $j f_0 \pm k 
f_{\mathrm m}$ with symmetrical amplitudes at the two sides.
This symmetry of the amplitudes can be seen from
the expression of amplitude ratio
$A(jf_0\pm kf_{\mathrm m})/A(jf_0)\sim 
\vert J_{\pm k}(ja^{\mathrm F}) \vert$,
and it is known that $J_{-k}(z)=(-1)^kJ_k(z)$.    
It is worth to compare the AM spectra 
in Figs.~\ref{am_sinF} and \ref{am_nsinF} to this FM spectrum.
The Fourier amplitude of the side peaks are proportional to 
the Bessel function, and an immediate consequence can be seen in the
figure: the amplitude of the triplet peaks are higher at $3f_0$ than
 at $2f_0$. (Although it is not shown in the figure,
the higher order $j > 5 $ harmonics have also smaller amplitudes 
than their side frequencies' ones.) 
Since the argument of the Bessel functions 
depends on the order of harmonics $j$, higher order harmonics 
``feel'' larger modulation index, which results in  more side peaks
around the higher order harmonics (see inserts in Fig.~\ref{fm_sinF}). 
This effect was found for V1127\,Aql from its {\it CoRoT\/} data
by \citet{Cha10}. 
A further remarkable flavour of this Fourier spectrum is, 
that it does not include  $f_{\mathrm m}$ as opposed to any 
of the AM spectra (insert in Fig.~\ref{fm_sinF}).

Let us return to the question whether there is any 
phase in the modulation cycle where the
light curve is identical to the
monoperiodic light curve (carrier signal). As 
it was shown in the case of sinusoidal AM modulation, 
there are some such possible phases.  Looking at the formula
(\ref{mod_fm_sin}) it can be realised that 
the modulation disappears at the moments of time if
$t=(l \pi - \varphi^{\mathrm F})/(2\pi f_{\mathrm m})$,
where $l$ is an arbitrary integer.

Estimating the number of the necessary parameters 
for a light curve fit the traditional description Eq.~(\ref{old}) needs  
$\approx 2n + 3+ 4 \sum_{j=1}^{n}[ {\mathrm {int}}(j a^{\mathrm F}) +1]$, 
where ``int'' means the integer function, and $n$ is the number of all
harmonics including the main frequency as well. At the same time,
Eq.~(\ref{mod_fm_sin}) requires only $2n+5$ parameters, no more than
in the sinusoidal AM case. For a typical 
case plotted in Fig.~\ref{Fig_fm_sin} ($n=10$ and
$a^{\mathrm F}=0.27$), the difference is 143 parameters
versus 25.

\subsubsection{The case of non-sinusoidal FM}\label{Sec_fm_nsin}

Assuming an arbitrary periodic modulation with a fixed frequency
we substitute a Fourier sum representing this modulation 
signal into Eq.~(\ref{mod_fm}) and get  
\begin{multline}\label{mod_fm_nsin}
m_{\mathrm {FM}}^*(t)= a_{0} + 
\sum_{j=1}^n{a_j \sin \Bigg[ 2\pi j f_0 t + {} } \\
{ \left. j \sum_{p=1}^{q}{
a^{\mathrm F}_p \sin \left( 2\pi p f_{\mathrm m} t 
+ \varphi^{\mathrm F}_p \right)} 
+ \phi_j \right]},
\end{multline}  
where the constant terms are contracted as
$\phi_j= j a^{\mathrm F}_0 + \varphi_j$.
As in the previous sinusoidal case the equation can be rewritten as
\begin{multline}\label{mod_fm_nsinF}
m^*_{\mathrm{FM}}(t)=  a_{0} +
\sum_{j=1}^n  
\sum_{k_1, k_2, \dots, k_p= -\infty}^{\infty}
a_j \left[ \prod_{p=1}^q J_{k_p}(ja^{\mathrm F}_p) \right]\cdot \\
\sin \left[  2\pi \left( j f_0 + \sum_{p=1}^q k_p p f_{\mathrm m} \right) t 
+ \sum_{p=1}^q  k_p \varphi_p^{\mathrm F} + \phi_j \right].
\end{multline}
In one sense, this formula is a generalisation of (\ref{mod_Sch}) 
to the case for a non-sinusoidal carrier wave, 
in the other sense however, 
the modulation frequencies are chosen specially as 
$f^p_{\mathrm m}:=p f_{\mathrm m}$.

\begin{figure} 
\includegraphics[width=9cm]{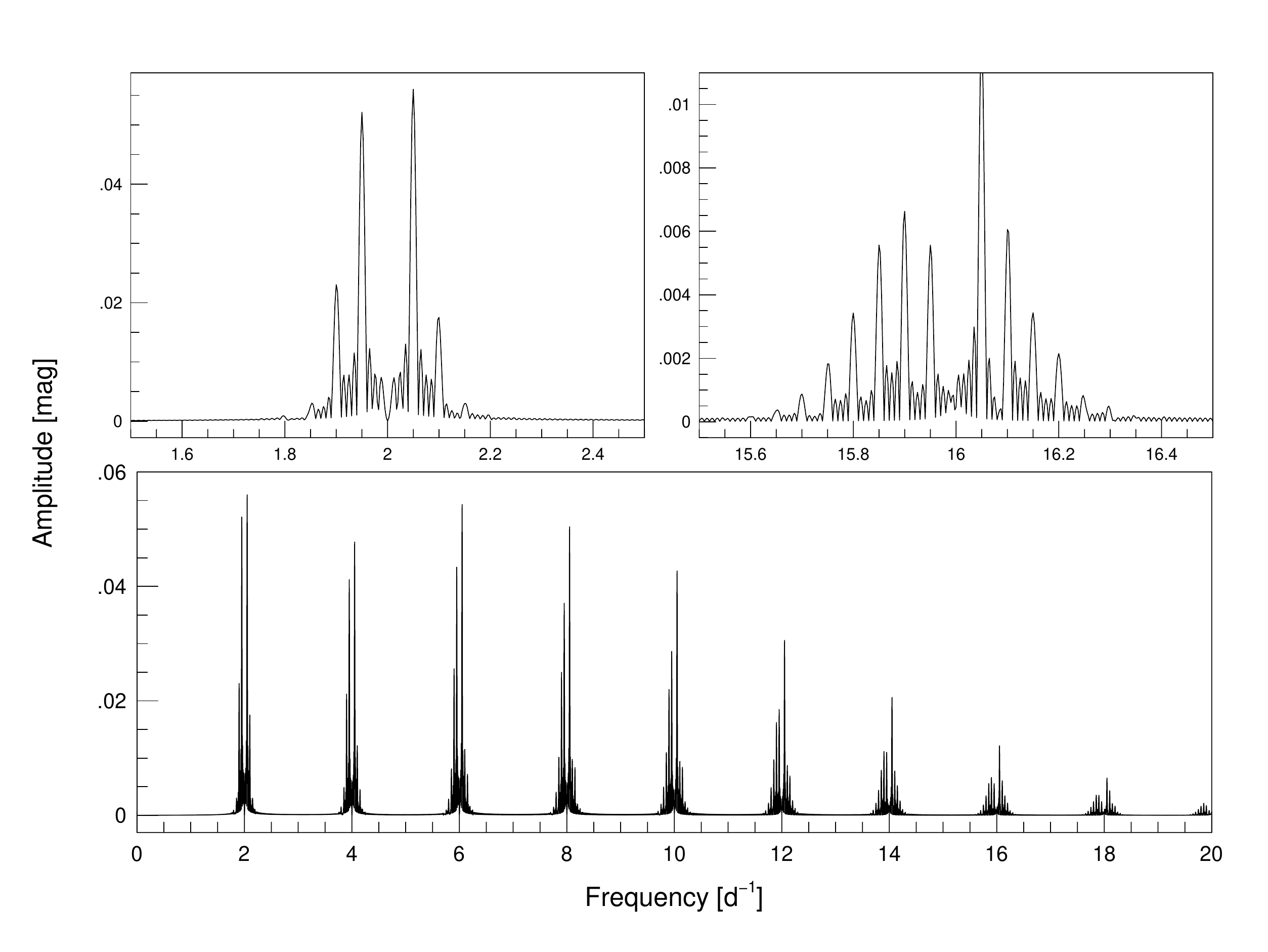}
\caption{Bottom:
Fourier amplitude spectrum of an artificial non-sinusoidal
FM light curve calculated by the formula of (\ref{mod_fm_nsin})   
after the data are prewhitened with the main frequency and its harmonics.  
Parameters of the generated light curve were the same as for
the light curve in Fig.~\ref{Fig_fm_sin}, 
and $p=2$, $a_2^{\mathrm F}=-0.1$~mag, 
$\varphi_2^{\mathrm F}=\pi /4$.
Top panels are zooms around the positions of the main 
frequency $f_0=2$~d$^{-1}$ (top left), and its 7th harmonic 
$8f_0=16$~d$^{-1}$ (top right), respectively.
}\label{fm_nsinF}
\end{figure} 

Comparing equation (\ref{mod_fm_nsinF}) with (\ref{mod_fm_sinF}) 
it can be realised that the structure
of both Fourier amplitude spectra should be similar
(cf. also Fig.~\ref{fm_sinF} and Fig.~\ref{fm_nsinF}),
although there are significant differences, as well.  First of all, 
besides the same values of common Fourier parameters 
of a sinusoidal and a non-sinusoidal case, the detectable
side peaks are more numerous than for the non-sinusoidal case.
The reason is simple: the higher order terms in the modulation
signals' sum increase the ``effective modulation index''. 
The most noteworthy difference is the disappearance of the symmetry between 
the amplitude of the side peaks from the lower and higher frequency parts. 

\begin{figure}
\includegraphics[width=9cm]{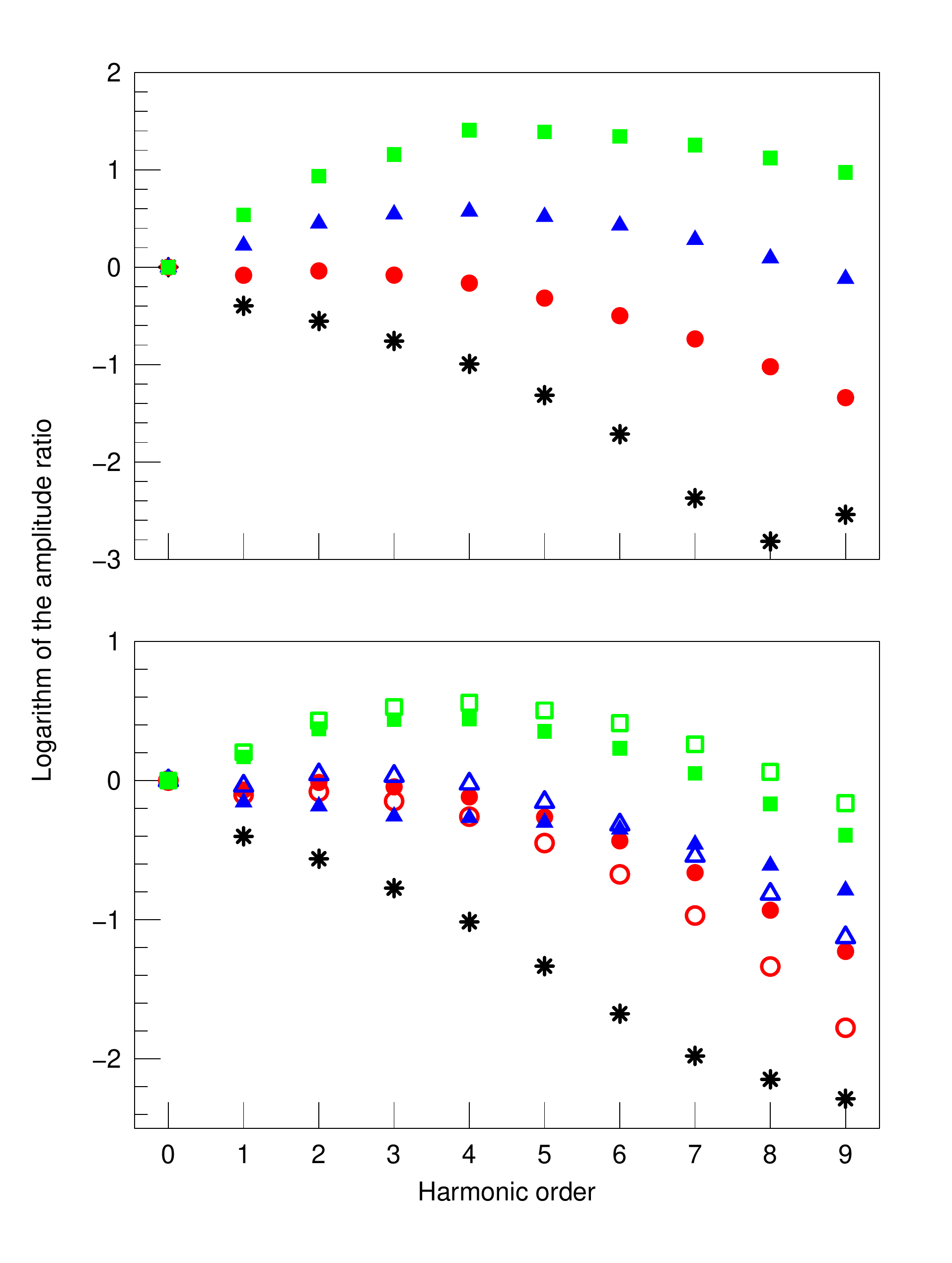}
\caption{
Amplitude ratios of the harmonic components of the main
pulsation ($A(jf_0)/A(f_0)$) 
plotted on decimal logarithmic scale compared
to the amplitude ratios of the modulation components:
$A(jf_0+pf_{\mathrm m})/A(f_0+pf_{\mathrm m})$. Top: sinusoidal 
FM case; $p=0$, 1, 2 and 3 (black) asterisks, (red) circles,
(blue) triangles and (green) squares, respectively.
Bottom: non-sinusoidal FM; shape of the symbols denotes $p$  
as in the top panel, but filled symbols mark positive $p$s (higher 
frequency side peaks) while the open symbols 
show negative $p$s (lower frequency side peaks).  
}\label{ampl_shape}
\end{figure}
\begin{figure}
\includegraphics[width=9cm]{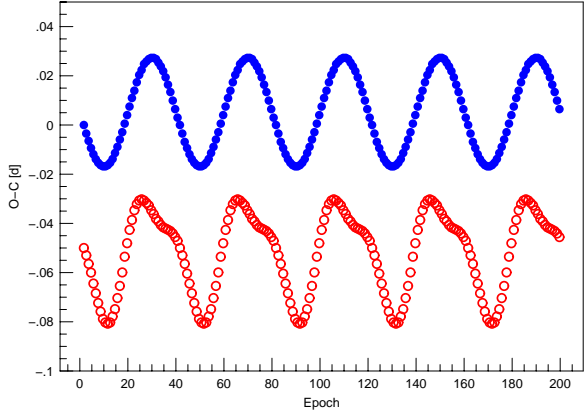}
\caption{
O-C diagram of the maxima of FM light curves
with sinusoidal -- (blue) filled circles -- 
and non-sinusoidal -- (red) empty circles -- modulations, respectively.
The input light curves are generated from the 
formulae of (\ref{mod_fm_sin}) and (\ref{mod_fm_nsin}) with the
same parameters as the light curve 
shown in Fig.~\ref{Fig_fm_sin} (sinusoidal 
case) and Fourier plot in Fig.~\ref{fm_nsinF} (non-sinusoidal case).
(For better visibility the non-sinusoidal curve is shifted by $-0.05$.)
}\label{o-c}
\end{figure}

To understand this let us 
investigate the simplest non-sinusoidal case, if $q=2$
and concentrate only on the side peaks around the main 
pulsation frequency ($j=1$). Then the 
above expression (\ref{mod_fm_nsinF}) is simplified to
\begin{multline}\label{nsin_fm2}
\sum_{k_1=-\infty}^{\infty} \sum_{k_2=-\infty}^{\infty}
a_1 J_{k_1}(a^{\mathrm F}_1) J_{k_2}(a^{\mathrm F}_2) \cdot \\
\sin \left\{ 2\pi \left[ f_0 + 
\left( k_1 + 2 k_2 \right)f_{\mathrm m} \right] t + 
k_1 \varphi^{\mathrm F}_1 + k_2 \varphi^{\mathrm F}_2 + \phi_1 \right\}.
\end{multline}
For calculating the amplitude of the triplet's side peaks 
$A(f_0\pm f_{\mathrm m})$ we have to sort out 
the corresponding terms from the above infinite sum 
such as $k_1=1-2k_2$ and $k_1=-(2k_2+1)$, for the 
right-hand-side and left-hand-side peaks, respectively 
($k_2$ is an arbitrary integer). It can be seen, that both  
sums include the same elements, because 
$J_{-3}(a^{\mathrm F}_1)J_{1}(a^{\mathrm F}_2)=
J_{3}(a^{\mathrm F}_1)J_{-1}(a^{\mathrm F}_2)$,
$J_{-5}(a^{\mathrm F}_1)J_{2}(a^{\mathrm F}_2)=
J_{5}(a^{\mathrm F}_1)J_{-2}(a^{\mathrm F}_2), \dots$ for 
each pair and the relative phase differences 
have the same values with an opposite sign.
The only differing terms contain 
$J_{1}(a^{\mathrm F}_1)J_{0}(a^{\mathrm F}_2)$ in 
the sum of  $A(f_0 + f_{\mathrm m})$ and 
$J_{-1}(a^{\mathrm F}_1)J_{0}(a^{\mathrm F}_2)$  in the 
sum of $A(f_0 - f_{\mathrm m})$. These terms are responsible
for the asymmetry of the side peaks. Introducing power 
difference of the side peaks as in Sec.~\ref{Basic_Bela}
we get 
\begin{equation}
\Delta_1=4 \hat{A}_1 J_{1}(a^{\mathrm F}_1)J_{0}(a^{\mathrm F}_2) 
\cos (\hat{\Phi}_1 - \varphi^{\mathrm F}_1). 
\end{equation}  
Here $\hat{A}_1$ and $\hat{\Phi}_1$ 
indicate the amplitude and phase of a sinusoidal 
oscillation obtained by summing all 
the terms in (\ref{nsin_fm2}) except the different ones.  
The asymmetry of the higher order side peaks
($\vert k_1+2k_2 \vert > 1$) can be verified in a similar manner.

This asymmetry has a further consequence.
The functions of amplitude ratio vs. harmonic orders belonging
to a given pair of side peaks
are diverge from each other (Fig.~\ref{ampl_shape}).
This behaviour is well-known from the similar
diagrams of observed Blazhko RR\,Lyrae stars \citep{JurDM,Cha10,Kol11}. 
It can also be seen that the actual character of the asymmetry
can change with the harmonic order $j$ or even within a 
given order with the different $p$. For example, in 
Fig.~\ref{ampl_shape} if $p=1$ (triplets), the right-hand-side
peaks are always higher than the left-hand-side ones and
the difference between the pairs are increasing with harmonic orders.
Meanwhile, if $p=3$ (septuplets) the situation is the
opposite. In the case of $p=2$ (quintuplets) the lower
frequency peaks have higher amplitude around the lower order ($j<5$) 
harmonics, for higher harmonics ($j>7$) the amplitude
ratios of the pairs of side peaks reversed.

As it was discussed in the introductory Section \ref{Basic_Bela}
 simultaneous and sinusoidal amplitude and phase modulations result 
in an asymmetrical spectrum, therefore, the asymmetry of the 
amplitude spectrum alone is not a good criterion for detecting a
non-sinusoidal FM.  The classical O$-$C 
diagram is an ideal tool for this purpose (see e.g. \citealt{Ste05}). 
Fig.~\ref{o-c} illustrates the O$-$C diagrams of the maxima 
for two artificial FM light curves: a sinusoidal and a simple
non-sinusoidal one. 

At the end of this section we compare the
necessary parameters of a potential fit based on the 
classical description (\ref{old}) and the present (\ref{mod_fm}) one.
In the latter case this value is $2n+2q+3$, where $n$ and
$q$ are defined in (\ref{mod_fm}). The expression is the
same as in the case of a non-sinusoidal AM. The traditional
formula needs $\approx 2n+3+ (4 \sum_{j=1}^{n} 
[{\mathrm{int}}(j \sum_{p=1}^{q} a^{\mathrm F}_p) +1 ] )$
parameters. For the case showed in Fig.~\ref{fm_nsinF} ($n=10$, 
$a^{\mathrm F}_1=0.27$, $a^{\mathrm F}_2=0.1$) these values are
27 and 163, respectively. 

\subsubsection{Parallel FM}

\begin{figure} 
\includegraphics[width=9cm]{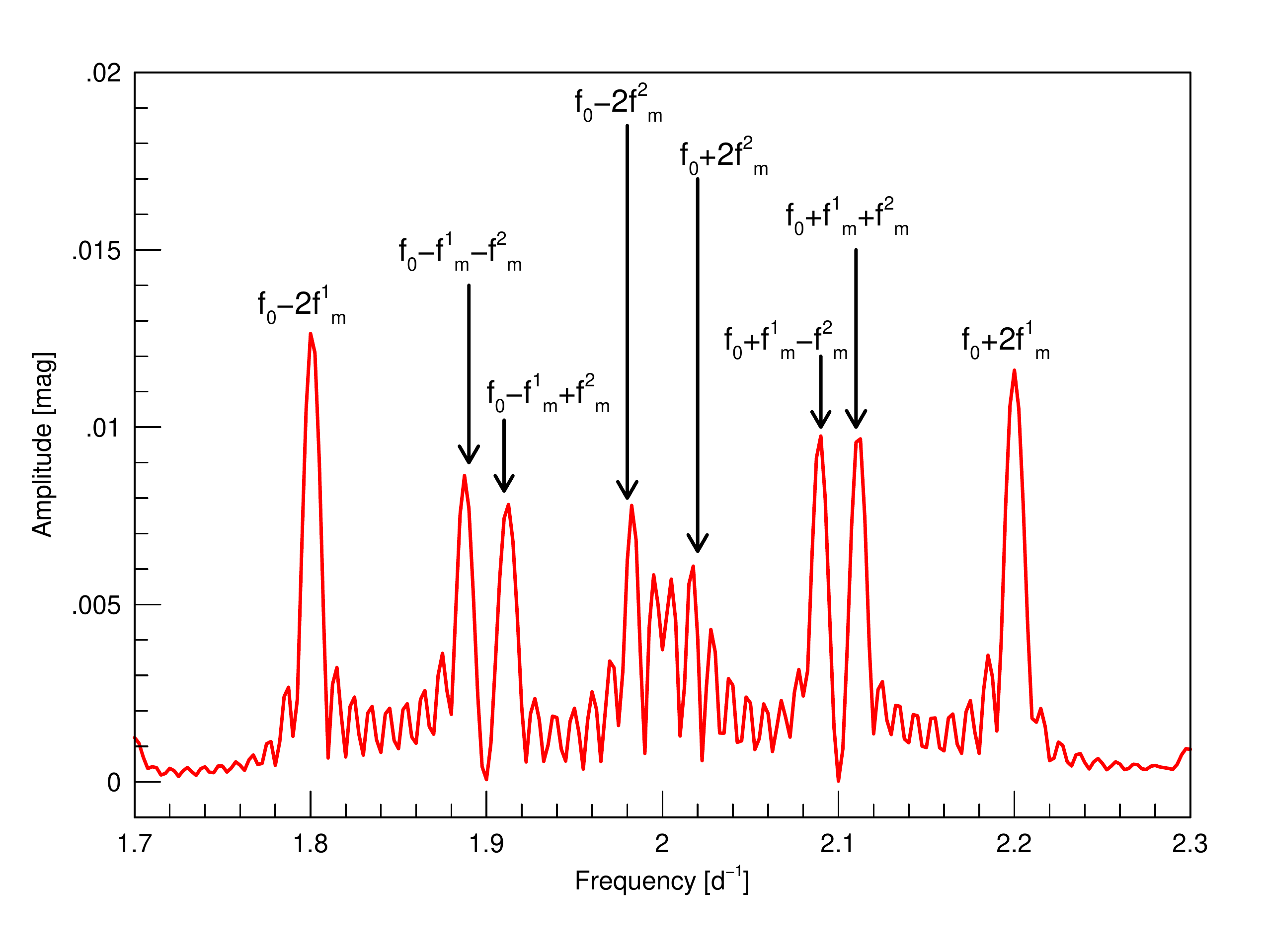}
\caption{Fourier amplitude spectra of the artificial light curves
containing two parallel FM modulations in formula (\ref{mod_fm_par}).
The figure shows a zoom around the main pulsation frequency after the data 
are prewhitened with it and the triplet components 
$f_0\pm \hat{f}^1_{\mathrm m}$, 
$f_0\pm \hat{f}^2_{\mathrm m}$. 
The highest peaks are at the quintuplet frequencies 
($f_0\pm 2\hat{f}^1_{\mathrm m}$, $f_0\pm 2\hat{f}^2_{\mathrm m}$) and 
at the linear combination frequencies. (The used parameters are:
$\hat{f}^1_{\mathrm m}=0.1$~d$^{-1}$, 
$\hat{f}^2_{\mathrm m}=0.01$~d$^{-1}$, 
$\hat{a}^{\mathrm F}_{11}=0.5$~mag, $\hat{a}^{\mathrm F}_{12}=0.2$~mag.)
}\label{fm_dp}
\end{figure}

We continue the discussion as in the case of AM.
The next step is the multiply modulated FM with independently 
superimposed modulation signals (parallel 
modulation). As has already been noted, the chance of such a 
scenario is very low for stars, but this case shows 
a new phenomenon, which is why it is worth to have a look at it. 
\begin{multline}\label{mod_fm_par} 
\hat{m}_{\mathrm {FM}}^*(t)= a_{0} + \sum_{j=1}^n a_j 
\sin \Bigg\{ 2\pi j f_0 t + {}  \\ 
\left. j \left[ \sum_{r=1}^s 
\sum_{p=1}^{q_r} \hat{a}^{\mathrm F}_{pr} 
\sin \left( 2\pi p \hat{f}^r_{\mathrm m} t + 
\hat{\varphi}^{\mathrm F}_{pr} \right) \right] + 
\hat{\phi}_j \right\}. 
\end{multline} 
Here $\hat{\phi}_j:=j \hat{a}^{\mathrm F}_{0r} + \varphi_j $.
The formula (\ref{mod_fm_par}) can easily be transformed with the help of 
Eqs.~(\ref{mod_Sch}) and (\ref{mod_fm_nsinF}) to 
\begin{multline}\label{mod_fm_parF} 
\hat{m}^*_{\mathrm{FM}}(t)= a_{0} + 
\sum_{j=1}^n 
\sum_{k_{11}, k_{12}, \dots, k_{{q_s} s}= -\infty}^{\infty} 
a_j \left[ \prod_{r=1}^{s} \prod_{p=1}^{q_r} 
J_{k_{pr}}(j \hat{a}^{\mathrm F}_{pr}) \right]\cdot \\ 
\sin \left[ 2\pi \left( j f_0 + \sum_{r=1}^{s} 
\sum_{p=1}^{q_r} k_{pr} p \hat{f}^r_{\mathrm m} \right) t + 
\sum_{r=1}^{s} \sum_{p=1}^{q_r} k_{pr} 
\hat{\varphi}_{pr}^{\mathrm F} + \hat{\phi}_j \right]. 
\end{multline} 
There is a fundamental difference between the construction of Fourier spectra 
of parallel AM and FM signals. While the AM spectra build up from a 
simple sum of the component spectra belonging to a given modulation 
frequency $\hat{f}^r_{\mathrm m}$, FM spectra contain all
possible linear combinations of  $\hat{f}^r_{\mathrm m}$ and the
harmonics of the main pulsation frequency $j f_0$. 
This is illustrated in Fig.~\ref{fm_dp}.
In practice, this effect complicates distinguishing 
Fourier spectra from the cascade AM and parallel FM.  

\subsubsection{The FM cascade}  

Although a parallel FM modulation results in a more complex
Fourier spectrum  than either a parallel or even a cascade AM, 
our former statement is still true. There is a very low chance  
for independently superimposed modulation signals in real stars.
Let us turn to the FM cascade (viz. the modulated modulation) case now!    
\begin{equation}\label{mod_fm_cas} 
\tilde{m}_{\mathrm {FM}}^*(t)= a_{0} + \sum_{j=1}^n a_j 
\sin \left[ 2\pi j f_0 t 
+ j \tilde{C}_{\mathrm {FM}}(t) + \tilde{\phi}_j  \right],
\end{equation} 
where
\begin{multline}\label{fm_casc}
\tilde{C}_{\mathrm {FM}}(t):=m_{\mathrm {FM}}^{(1)}(t)=
\sum_{p=1}^{q_1}\tilde{a}^{\mathrm F}_{p1} 
\sin \left[ 2\pi p \tilde{f}^1_{\mathrm m} 
t + p m_{\mathrm {FM}}^{(2)}(t) + \tilde{\varphi}^{\mathrm F}_{p1} \right], \\
m_{\mathrm {FM}}^{(2)}(t)=
\sum_{p=1}^{q_2}\tilde{a}^{\mathrm F}_{p2} 
\sin \left[ 2\pi p \tilde{f}^2_{\mathrm m}
t +  p m_{\mathrm {FM}}^{(3)}(t) + \tilde{\varphi}^{\mathrm F}_{p2} \right], \dots, \\
m_{\mathrm {FM}}^{(s)}(t)=
\sum_{p=1}^{q_s}\tilde{a}^{\mathrm F}_{ps} 
\sin \left( 2\pi p \tilde{f}^s_{\mathrm m}
t + \tilde{\varphi}^{\mathrm F}_{ps} \right),
\end{multline}
and $\tilde{\phi}_j=j \tilde{a}^{\mathrm F}_{01}+\varphi_j$.
As we can see, the function $\tilde{C}_{\mathrm {FM}}(t)$
consists of a modulation cascade with $s$ elements, where all
elemental modulation functions $m_{\mathrm {FM}}^{(r)}(t)$ are 
represented by finite Fourier sums. That is, they are assumed to be
independent periodic signals with the frequencies 
$\tilde{f}^r_{\mathrm m}$. Since an FM modulation
can be reproduced by infinite series of 
sinusoidal functions (see Chowning relation),
it is not a surprise that the sinusoidal decomposition of 
the expression (\ref{mod_fm_cas}) 
is very similar to the parallel case (\ref{mod_fm_parF}).
Namely
\begin{multline}\label{mod_fm_casF} 
\tilde{m}^*_{\mathrm{FM}}(t)= a_{0} + 
\sum_{j=1}^n 
\sum_{k_{11}, k_{12}, \dots, k_{{q_s} s}= -\infty}^{\infty} a_j \cdot \\ 
\left[ \prod_{p=1}^{q_1} J_{k_{p1}}\left( j \tilde{a}^{\mathrm F}_{p1} \right)
\prod_{r=2}^{s} \prod_{p=1}^{q_r} 
J_{k_{pr}}\left( k_{p-1,r} \tilde{a}^{\mathrm F}_{pr} \right) \right] \cdot \\ 
\sin \left[ 2\pi \left( j f_0 + \sum_{r=1}^{s} 
\sum_{p=1}^{q_r} k_{pr} p \tilde{f}^r_{\mathrm m} \right) t + 
\sum_{r=1}^{s} \sum_{p=1}^{q_r} k_{pr} 
\tilde{\varphi}_{pr}^{\mathrm F} + \tilde{\phi}_j \right]. 
\end{multline} 
The frequency content is exactly the same as in the parallel case, only
the values of amplitudes and phases are different.

\subsection{The case of PM}

Here we discuss the phase modulation. As we stated in the
Sec.~\ref{Basic} there is no chance to distinguish between 
FM and PM phenomena on the basis of their measured signals (inverse problem), 
if the modulation function $m^*_{\mathrm m}(t)$ in Eq.~(\ref{mod_fm}) 
is allowed to be arbitrary. At the same time, if the basic physical 
parameters such as effective 
temperature, radius and $\log g$ are changing during the Blazhko cycle
as was found recently \citep{Sod09, JurMWII, JurDM} the cyclic
variation of the fundamental pulsation period (vis. frequency)
that results in FM would be a plausible explanation for observed
effects. There is an additional possible
argument against the existence of PM in RR\,Lyrae stars.

If we assume that the modulating function 
$m^*_{\mathrm m}$ contains no explicit time variation --
as in the usual definition for PM modulation in electronics -- 
Eq.~(\ref{mod_fm}) reads as
\begin{equation}\label{mod_pm}
  m_{\mathrm{PM}}^*(t)= a_{0} + 
\sum_{j=1}^n{a_j 
\sin \left[ 2\pi j f_0 t + m^*_{\mathrm m}(t) + \varphi_j \right] }.
\end{equation} 
When this formula is expressed as Eq.~(\ref{mod_fm_sinF}) or
Eq.~(\ref{mod_fm_nsinF}) according to a sinusoidal or an arbitrary 
periodic modulating function, respectively, the arguments of Bessel
functions are independent from the harmonic order $j$ as opposed
to the case of FM. It causes a systematic difference between 
Fourier spectra of FM and PM. While the number of detectable 
side peaks in FM increases with the order of harmonics, 
for PM the number of side peaks is the same for all 
harmonics. 

There are two  Blazhko RR\,Lyrae stars that show both 
strong phase variations and their data are precise enough, 
these are the {\it CoRoT\/} targets V1127\,Aql
and CoRoT~105288363 \citep{Cha10, Gug11}.
The spectrum of V1127\,Aql clearly shows the existence 
of FM: 3rd order side frequencies are detected
 around the main pulsation frequency while order of 8th around
the 19th harmonic. The Fourier analysis of separate Blazhko 
cycles of CoRoT~105288363 showed that with the increasing strength
of phase variation, the number of detected side peaks around higher
order harmonics are also increased \citep{Gug11}. 
It is an evidence of (changing) FM.

\subsection{Real Blazhko stars with simultaneous AM and FM}

In this section we discuss the general combined
case, when both types of modulations occur simultaneously.
As it was mentioned before both AM and FM type modulations 
were detected for all observed Blazhko RR\,Lyrae stars 
if the observed data sets were precise and long enough.
This is the situation for ground-based 
(\citealt{Jur09}; and references therein) and 
space-born observations of {\it CoRoT\/} and {\it Kepler\/} 
\citep{Cha10, Por10, Ben10, Kol11} as well.  

Generalising the sinusoidal case of combined modulation
Eq.~(\ref{def_comb}) discussed in Sec.~\ref{Basic_Bela} we get
\begin{equation}\label{Comb_general}
m^*_{\mathrm {Comb}}(t)=\left[ 1 + m^*_{\mathrm m}(t) \right] 
m^*_{\mathrm {FM}}(t),
\end{equation}
where $m^*_{\mathrm {FM}}(t)$ is the general modulated FM function
defined by Eq.~(\ref{mod_fm}). Since all observed Blazhko stars
show AM and FM with the same frequency, we have investigated only those
cases where this assumption is fulfilled.

\subsubsection{Combined modulations with sinusoidal 
functions}\label{combsin}

The simplest case similarly to the pure AM and FM cases is the
simultaneous but sinusoidal modulations.
\begin{multline}\label{mod_comb_sin}
  m^*_{\mathrm{Comb}}(t)=
\left( 1 + h \sin 2\pi f_{\mathrm m}t \right) \cdot \\
\left\{ a_0 + \sum_{j=1}^n a_j 
\sin \left[ 2\pi j f_0 t + j a^{\mathrm F} 
\sin \left( 2\pi f_{\mathrm m} t + \phi_{\mathrm m} \right) 
+ \varphi_j \right] \right\},
\end{multline}
where the notations are the same or directly analogous with the
previously defined ones: $h=a_{\mathrm m}/U^*_{\mathrm {FM}}$
and $U^*_{\mathrm {FM}}$ is the amplitude of the second term 
(the FM modulated ``carrier wave''). The relative phase
between AM and FM signals is 
$\phi_{\mathrm m}=\varphi^{\mathrm F} - \varphi_{\mathrm m}$.

\begin{figure} 
\includegraphics[width=9cm]{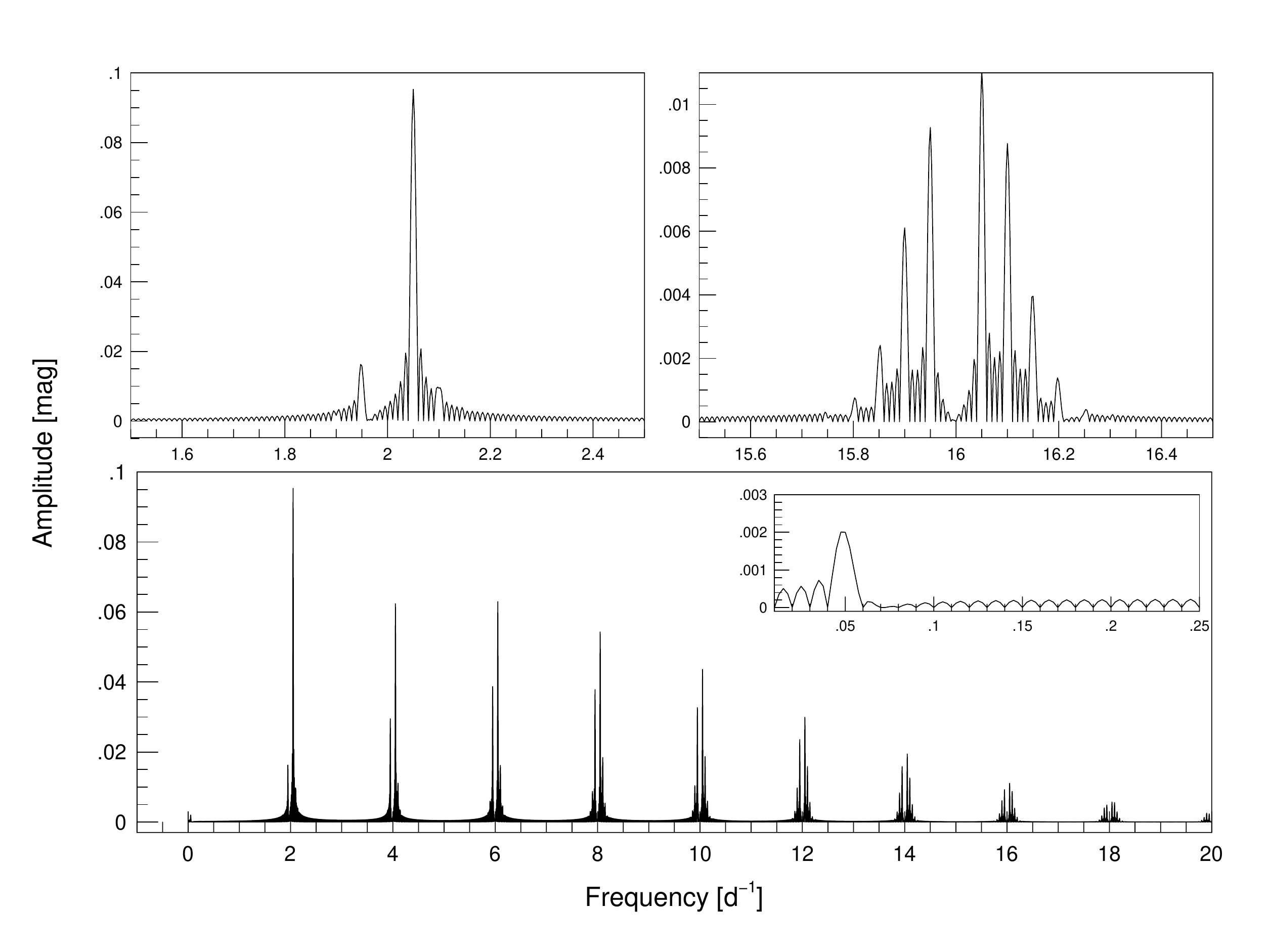}
\caption{Bottom:
Fourier amplitude spectrum of an artificial combined
(AM \& FM) light curve computed
from  Eq.~(\ref{mod_comb_sin}) after the data 
are prewhitened with the main frequency and its harmonics.  
Top panels are zooms around the positions of the main 
frequency $f_0=2$~d$^{-1}$ (top left), and its 7th harmonics 
$8f_0=16$~d$^{-1}$ (top right), respectively.
The relative phase between AM and FM is set to 
$\phi_{\mathrm m}=270$~deg.
}\label{comb_sinF}
\end{figure}

According to the schema of (\ref{mod_Comb})
expression (\ref{mod_comb_sin}) can be reformulated into
\begin{multline}\label{mod_comb_sinF}
 m^*_{\mathrm{Comb}}(t)= a_0 + a_0 h \sin 2\pi f_{\mathrm m}t + \\ 
\sum_{j=1}^{n}
\sum^{\infty}_{k=-\infty} a_j \left\{ J_k(j a^{\mathrm F}) 
\sin \left[ 2 \pi \left( f_0 + 
kf_{\mathrm m} \right) t + k\phi_{\mathrm m} + 
\varphi_j \right] + \right. {} \\
\left. \frac{h}{2}J_{k-1}(j a^{\mathrm F}) 
\sin \left[ 2 \pi \left( f_0 + kf_{\mathrm m} \right) t + 
\left( k-1 \right) \phi_{\mathrm m} + \varphi^{-}_j \right] + \right. {} \\ 
\left. \frac{h}{2}J_{k+1}(j a^{\mathrm F} ) 
\sin \left[ 2 \pi \left( f_0 + kf_{\mathrm m} \right) t + 
\left( k+1 \right) \phi_{\mathrm m} + \varphi^{+}_j \right] \right\},
\end{multline}
where $\varphi^{\pm}_j=\varphi_j\pm \pi /2$.
Based on Eq.~(\ref{mod_comb_sin})
the Fourier spectrum in Fig.~\ref{comb_sinF}  
can be interpreted as a sum of
the combined modulation with sinusoidal carrier wave 
(\ref{mod_Comb}) with an additional term describing
the modulation frequency itself (insert in bottom panel).
Each harmonic is surrounded by a multiplet
structure of peaks just like the main frequency. The number of
side peaks increases with the harmonic order $j$ 
similarly for FM (Sec.~\ref{Sec_sinfm}).

The asymmetrical amplitudes of pairs of side frequencies 
belonging to a given harmonic can be characterised similarly
to the sinusoidal carrier wave case (\ref{asymm}) as
\begin{equation}\label{asym_sin}
\Delta_{jl}=-4\frac{h l}{j a^{\mathrm F}} 
a^2_j J^2_l(j a^{\mathrm F}) \sin \phi_{\mathrm m}.
\end{equation}
Here $\Delta_{jl}=A^2(jf_0+l f_{\mathrm m}) - 
A^2(jf_0 - l f_{\mathrm m})$ is the power difference
of the $l$th side peaks at the $j$th harmonics ($l=1,2,\dots$).
Similarly to the course book case discussed in Sec.~\ref{Basic_Bela}
the asymmetry depends on the actual value of $h$ and $a^{\mathrm F}$,
(viz. the relative strengths of AM and FM) and the
relative initial phase angle $\phi_{\mathrm m}$. 
The most extreme possibility is when one of the side
peaks completely disappears. 
The necessary conditions are $\phi_m=\pm \pi /2$ and 
$ja^{\mathrm F}=h l$.
The asymmetry decreases with the increasing harmonic order $j$ 
(see also top panels in Fig.~\ref{comb_sinF}), because all the Bessel 
functions quickly converge to zero with increasing arguments,
therefore dominate the right hand side of expression (\ref{asym_sin}).

Non-equidistant sampling and large gaps 
in the observed time series can cause significant differences 
between side-peak amplitudes (see \citealt{JurRR}). 
Such sampling effects, however, can not explain huge differences,
such as when side peaks completely disappear in one side
and the spectra show doublets, though numerous examples were found  
by large surveys as MACHO and OGLE \citep{Alc00, Alc03, MP03}.
But as  illustrated by  Fig.~\ref{comb_sinF}, 
highly asymmetrical side peaks can easily be generated by 
(\ref{mod_comb_sin}). This asymmetry effect 
can be a possible explanation for the observed 
doublets (RR-$\nu$1 stars) and even for triplets (RR-$\nu$2 stars). 
In the latter 
case the two side frequencies can originate from a quintuplet
structure (equidistant triplet on one side) or from 
a multifrequency modulation (non-equidistant triplet on one side).

\begin{figure} 
\includegraphics[width=9cm]{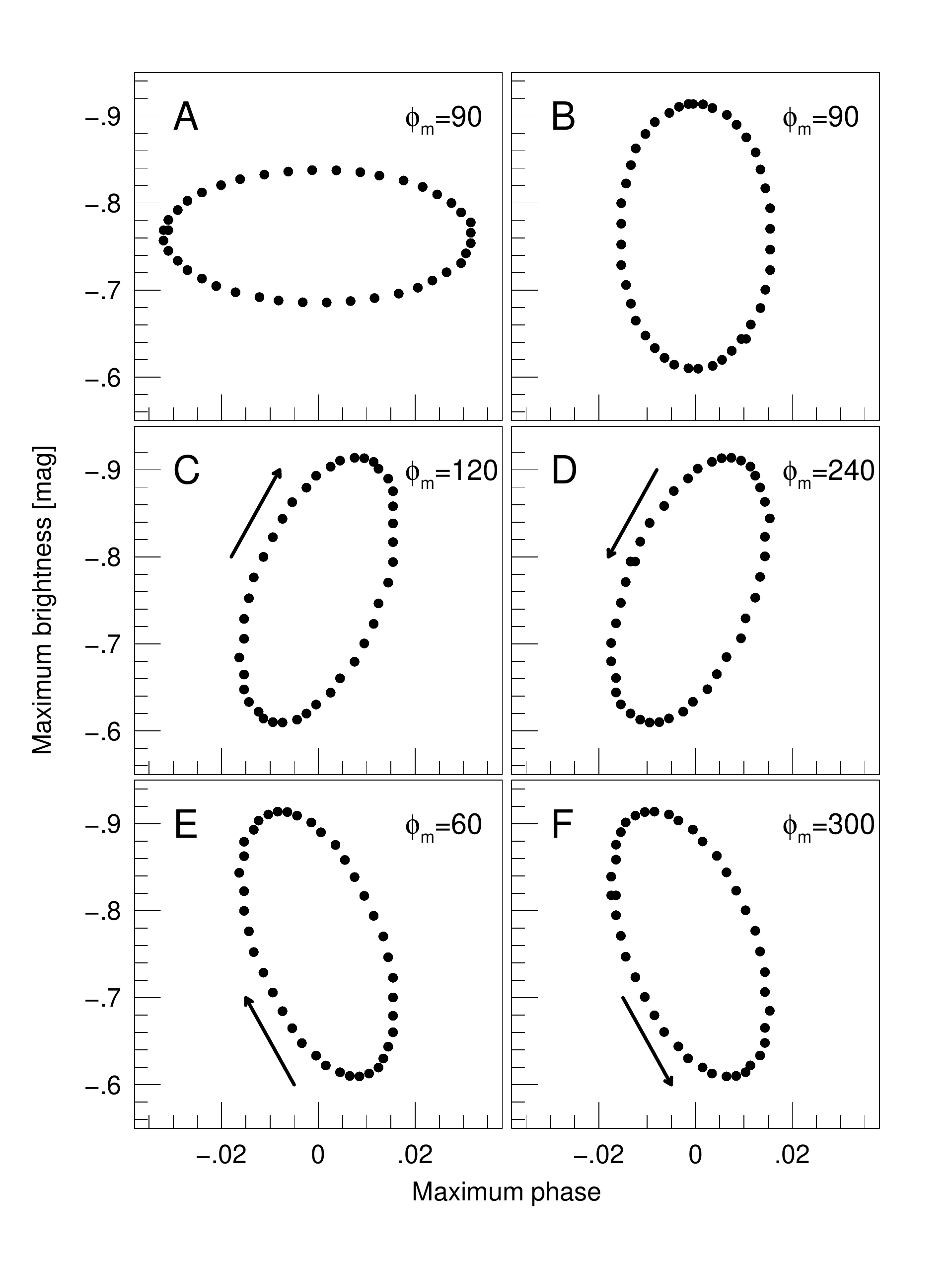}
\caption{
Maximum brightness vs. maximum phase diagrams for
some artificial light curves with combined sinusoidal
modulations. Between panels A and B the relative strengths
of AM and FM are changed as (A) $h=0.1$ and $a^{\mathrm F}=0.2$
and (B) $h=0.2$ and $a^{\mathrm F}=0.1$, respectively.
From panel B to F the amplitudes are fixed and only the
relative phase $\phi_{\mathrm m}$ is changed as shown 
at the upper left corner in each panel. 
The arrows in panels C-F indicate the direction of motion.
}\label{egg_sin}
\end{figure}

Searching for phases where the 
modulated and non-modulated light curves are identical
we conclude that such phases exist only if $\phi_m=(k_1-k_2) \pi$;
($k_1$, $k_2$ are integers) and then 
the moments of the coincidences are $t= k_2 /2 f_{\mathrm m}$.
Assuming  all Blazhko stars showing both AM and FM 
this conclusion supports the finding of \cite{Jur02}, who studied light 
and radial velocity curves of Blazhko RR\,Lyrae stars,

The amplitude ratio vs. harmonic order
diagrams show similar shapes and relative positions as were discussed
in Sec.~\ref{Sec_fm_nsin} with the connection of Fig.~\ref{ampl_shape}. 
Let us look at the maximum brightness vs. maximum phase diagrams,
 a classical tool for analysing Blazhko RR\,Lyrae stars.
Such diagrams are plotted in Fig.~\ref{egg_sin} 
for synthetic light curves generated from the formula 
of (\ref{mod_comb_sin}).
All diagrams have a simple round shape. 
They reflect the relative strength of AM and FM components.  
In panels A and B the relative strengths are 
opposite $2h=a^{\mathrm F}$ and $h=2a^{\mathrm F}$, respectively.
As a consequence, the loop is deformed vertically or horizontally.  
When the angle $\phi_{\mathrm m}$ differs from
the special values of $l \pi /2$, ($l=0,1, 2, 3, 4$),
the axes of the loops 
are inclined to the vertical horizontal position.
This angle also determines the direction of motion.
If $0< \phi_{\mathrm m} < \pi$ it is clockwise, 
whilst if $\pi< \phi_{\mathrm m} < 2\pi$ it is anti-clockwise.
(These conditions are the same as it was found by \citealt{Sz09}
for sinusoidal carrier waves.)
It is noteworthy, that the same ranges of $\phi_{\mathrm m}$
also determine the character of power difference of the side peaks:
if the right hand side peaks are higher than 
the left hand side ones then the 
direction of motion is anti-clockwise and vice versa. 

\subsubsection{Non-sinusoidal combined modulation}

On the basis of the previous sections 
it is easy to define the light curves
which are modulated by general periodic signals 
simultaneously both in their amplitudes and phases: 
\begin{equation}
m^*_{\mathrm{Comb}}(t)= 
\frac{m^*_{\mathrm{AM}}(t)}{c^*(t)} m^*_{\mathrm{FM}}(t),
\end{equation}
where the functions $m^*_{\mathrm{AM}}(t)$ and 
$m^*_{\mathrm{FM}}(t)$ are defined by Eqs.~(\ref{mod_AM}) and 
(\ref{mod_fm_nsin}), respectively. Since the 
two modulations are described by different functions 
(they are represented by different order of Fourier sums), 
no such simple relative phase can be defined as 
$\phi_{\mathrm m}$ for the sinusoidal case in Sec.\ref{combsin}.
Therefore, we obtain for the mathematical form of such a 
generally modulated light curve:  
\begin{multline}\label{mod_nsinC}
m_{\mathrm{Comb}}^*(t)=\left[ a^{\mathrm A}_0+ 
\sum_{p'=1}^{q'} a^{\mathrm A}_{p'} 
\sin\left( 2\pi p' f_{\mathrm m} t + 
\varphi^{\mathrm A}_{p'} \right) \right] \cdot \\
\left\{ a_{0} + 
\sum_{j=1}^n a_j \sin \Bigg[ 2\pi j f_0 t + \right. \\
\left. \left. j \sum_{p=1}^{q} a^{\mathrm F}_p 
\sin \left( 2\pi p f_{\mathrm m} t + \varphi^{\mathrm F}_p \right) 
+ \phi_j \right] \right\},
\end{multline}  
where the notations are the same as in Eqs.~(\ref{mod_AM}) 
and (\ref{mod_fm_nsin}). This expression describes
all the discussed phenomena of a light curve
modulated regularly with a single frequency $f_{\mathrm m}$.
The envelopes of these light curves are very similar to 
the envelopes of non-sinusoidal AM light curves shown in
Fig.~\ref{am_nsinlc}. The light curves show 
non-sinusoidal phase variation as well (see also Figs.~\ref{Fig_fm_sin}
and \ref{o-c}). 

As in the former simpler cases,
the Fourier spectrum can also be constructed 
analytically with the help of the sinusoidal decomposition
of (\ref{mod_nsinC}): 
\begin{multline}\label{mod_comb_nsinF}
m^*_{\mathrm{Comb}}(t)=  a_{0} a^{\mathrm A}_0+
\sum_{p'=1}^{q'}a_{0} a^{\mathrm A}_{p'} 
\sin \left(2 \pi p' f_{\mathrm m} t + \varphi^{\mathrm A}_{p'} \right) + \\ 
\sum_{j=1}^n  \sum_{p'= 0}^{q'} 
\sum_{k_1, k_2, \dots, k_q= -\infty}^{\infty} 
\frac{a^{\mathrm A}_{p'}}{2} a_j 
\left[ \prod_{p=1}^q J_{k_p}(ja^{\mathrm F}_p) \right] \cdot \\
\sin \left\{  2\pi \left[ j f_0 + \left(\sum_{p=1}^q k_p p \pm p'\right)
f_{\mathrm m} \right] t 
+ \psi^{\pm}_{pp'j} \right\}.
\end{multline}
Here the 
$\psi^{\pm}_{pp'j}:= \sum_{p=1}^{q}{k_p \varphi^{\mathrm F}_p}
\pm \varphi^{\mathrm A}_{p'} + \phi_j \mp \pi/2$, the arbitrary 
constant is chosen as $\varphi^{\mathrm A}_{0}:=\pi /2$.
The qualitative structure of this spectrum is simple
and well-understandable on the basis of the previously discussed cases.
The second term is responsible for the appearance of the modulation 
frequency and its higher harmonics (see insert in Fig.~\ref{am_nsinF}). 
The next (infinite number of) terms
describe a spectrum which is similar to the non-sinusoidal
FM spectrum (Fig.~\ref{fm_nsinF}) but it also shows
the AM splitting which is present in the sinusoidal combined case. 
These effects make the calculation of the peaks' amplitude complicated.
The asymmetry between each pair from a multiplet 
around a given harmonic is determined by two factors:
 one of them is the the non-sinusoidal nature of the FM  
(Sec.~\ref{Sec_fm_nsin}) and 
the other one is the combination of AM and FM (Sec.~\ref{combsin}). 

\begin{figure} 
\includegraphics[width=9cm]{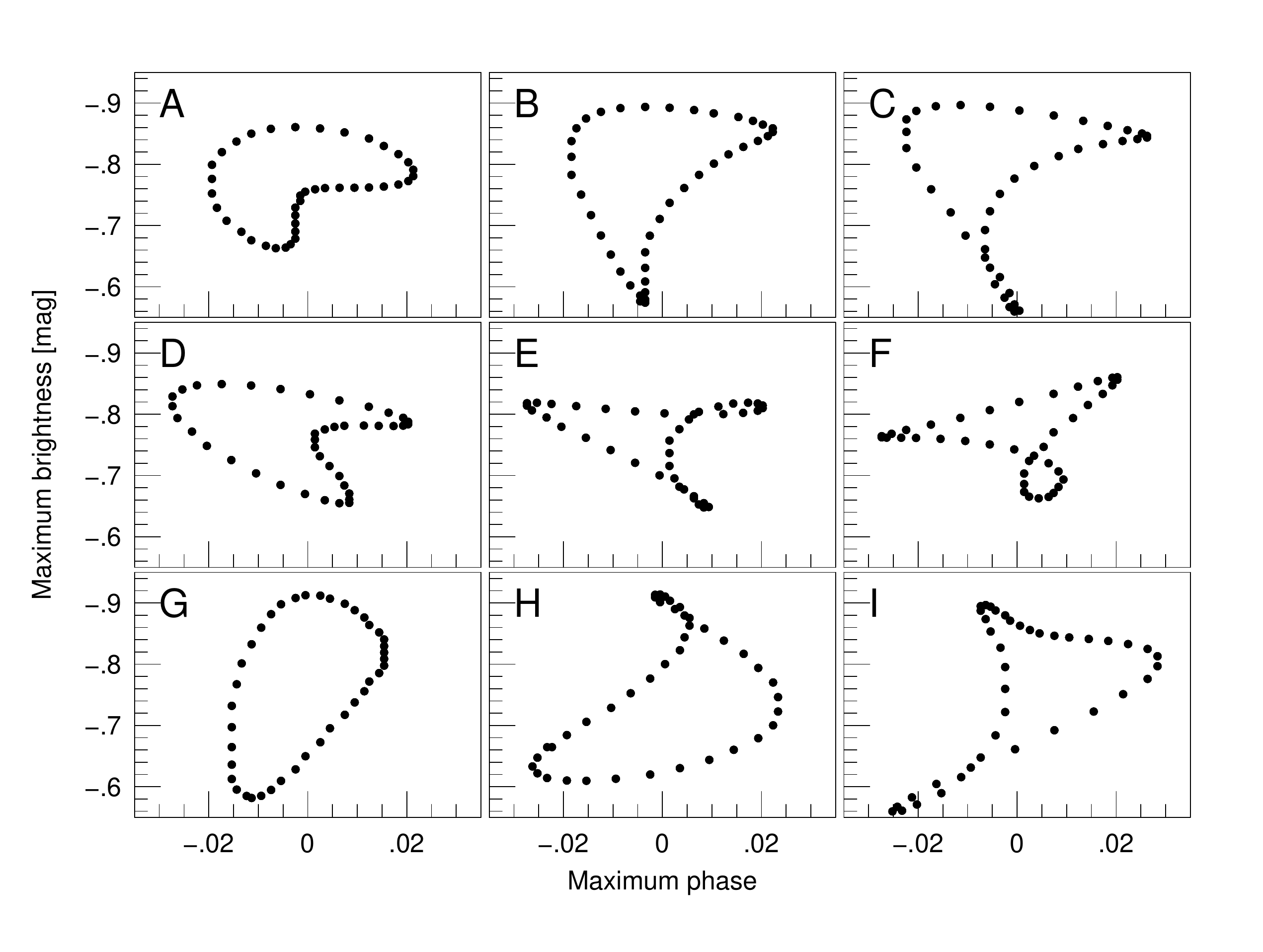}
\caption{
Some typical maximum brightness vs. maximum phase diagrams for
synthetic light curves with combined non-sinusoidal
modulations. The relative strength, initial phases and
number of used harmonics of AM and FM has been varied.
}\label{egg_nsin}
\end{figure}

The maximum  brightness vs. maximum phase diagrams generally
show complicated shapes. They could form knots, loops and
other non-trivial features. A collection of such diagrams 
is plotted in Fig.~\ref{egg_nsin}. The direction of the motion
can arbitrarily change by tuning the initial phases. 

\subsubsection{Combined multifrequency modulations}

A combined modulation with multiperiodic AM or FM or both
can be handled analogously to the simpler presented cases.
We can substitute $m^*_{\mathrm m}(t)$ into the general
expression (\ref{Comb_general}) as it was defined by Eq.~(\ref{mod_MAM})
(parallel AM) or Eq.~(\ref{mod_MAM2}) (AM cascade). 
Writing $m^*_{\mathrm {FM}}(t)$ as
 Eq.~(\ref{mod_fm_par}) (parallel FM) or
Eq.~(\ref{mod_fm_cas}) (FM cascade) in principle is straightforward.
In practice, however, calculating  
coefficients (amplitudes, phases) is more complicated. 
The resulting light curves and Fourier spectra can be 
interpreted on the basis of their constituents.
They do not show new features except their  
maximum brightness vs. maximum phase diagrams which
show time-dependent and generally non-closed
curves as opposed to those in Fig~\ref{egg_nsin}. 
If the ratio of modulating frequencies are commensurable,
the curve is closed, otherwise it has
a non-repetitive behaviour.  
The reason is that if the modulation is described by 
$N$ independent frequencies the proper diagram would be 2$N$-dimensional
and the classical one is only a 2-D projection of it.

\section{Practical application -- a case study}

\begin{figure*} 
\includegraphics[width=18cm]{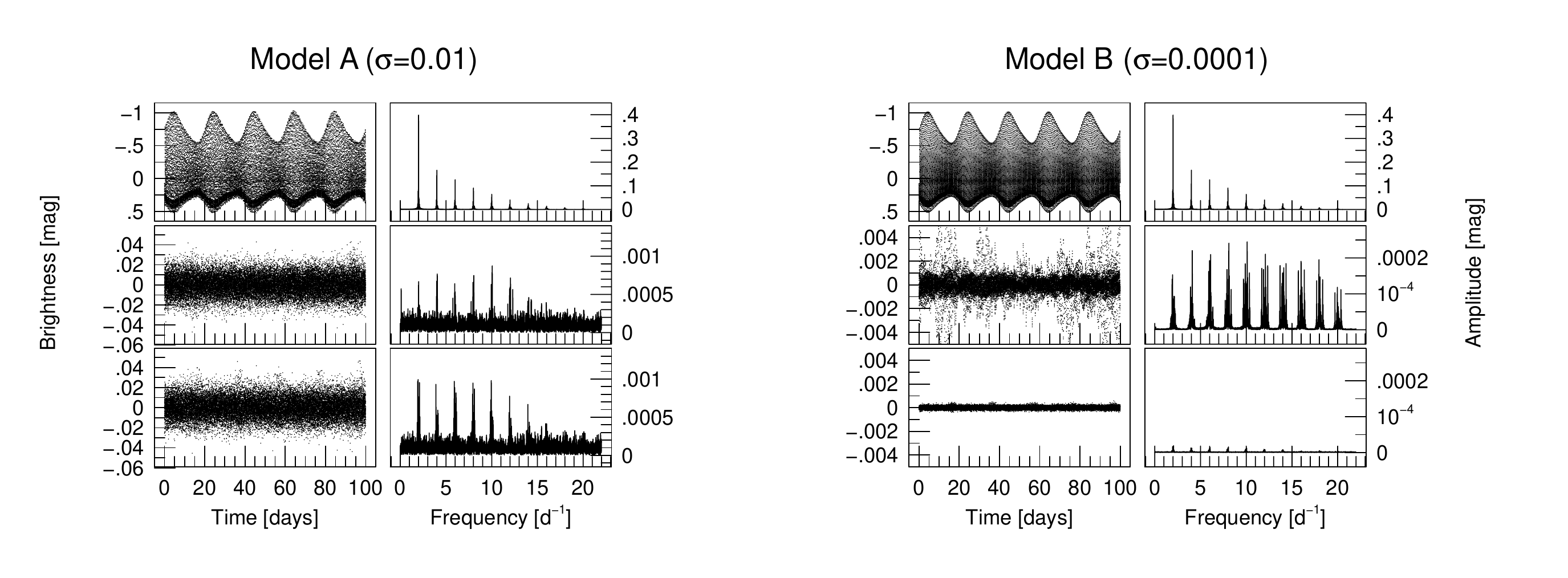}
\caption{
Artificial light curves 
with Gaussian noise and their Fourier spectra.
Model A ($\sigma=10^{-2}$) is plotted in the left,
model B ($\sigma=10^{-4}$) in the right.
The residual light curves and their spectra after the 
traditional fitting process (middle panels) and 
the present one (bottom panels). (Residuals and their
spectra are in different scales for the two models.)
}\label{appl_Fig}
\end{figure*}

To demonstrate how our formalism works in practice,
we generated two artificial light curves with simultaneous
non-sinusoidal AM (with two harmonics, $p'=2$) and FM 
(with three harmonics, $p=3$). 
The light curves are 100-days long
and sampled by 5-minutes in the same manner as the all 
synthetic light curves in the paper. 
We added Gaussian noise to
the light curves either with rms=0.01~mag  (model A) or $10^{-4}$~mag 
(model B), respectively. 
The model A is similar to a good quality ground-based observation,
while the model B simulates a typical space-borne data set.
These two artificial light curves (top panels in Fig.~\ref{appl_Fig})
were analysed with a blind test (i.e. without 
knowledge about frequencies, amplitudes and phases) 
both by the traditional way and by our method. 

\subsection{Classical light curve analysis}

Constructing the mathematical model Eq.~(\ref{old}) of the light curves
in the traditional analysis is a successive prewhitening process.
It consists of Fourier spectra building, 
fitting the data with the parameters of 
the highest peak(s) in the spectrum by a non-linear algorithm,
and subtracting the fitted function from the data and so on.
The process continues as long as significant
peaks are detected. At the end of this analysis the noise
of the residual data reaches the observational scatter.
 
In the case of model A the highest peaks belong to
the main frequency $f_0$ and its harmonics. In the further
prewhitening steps the triplets ($f_0\pm f_{\mathrm m}$),
quintuplets ($f_0\pm 2f_{\mathrm m}$), septuplets 
($f_0\pm 3f_{\mathrm m}$), nonuplets ($f_0\pm 4f_{\mathrm m}$)
and the modulation frequency ($f_{\mathrm m}$) were found and fitted.
The significance level was chosen at $S/N=4$, where  
signal-to-noise ratio ($S/N$) is estimated as \cite{Bre93}.
To remove all significant peaks from the spectrum
five from the undecaplet peaks ($f_0+ 5f_{\mathrm m}$) and
two aliases have to be fitted and subtracted.
(We note that the frequency $2f_{\mathrm m}$ was detectable, 
but under the significance level, therefore it was not fitted.)
The residual light curve of this process and its Fourier spectrum 
are plotted in the middle panels of Fig.~\ref{appl_Fig}. 
The rms of the residual light curve is $10^{-2}$, so
we got back the input noise value. The number of 
fitted frequencies are 103 and used parameters
in this successive fit is 201. If the frequency of
the side peaks are also fitted independently 
(``let it free approach'') this value increases to 286.

The process works similarly in the case of model B as well.
Naturally, many more significant peaks are detectable.
From the highest peaks to the lowest:
$f_0$, its harmonics, the side peaks up their orders of
six ($f_0\pm 6f_{\mathrm m}$), the modulation frequency
and its harmonic ($f_{\mathrm m}$, $2f_{\mathrm m}$) are significant.
Many alias peaks originating from 
the finite data length are also detectable:
46 such frequencies were removed up to the significance
level of the sixth order of side peaks. 
We stopped the analysis here at $S/N\approx{40}$,
because we already found 168 frequencies and used 368 parameters
(or 478 if we fit each frequency independently).
The resulted light curve and its Fourier spectrum are
shown in the middle panels of Fig.~\ref{appl_Fig}.
The rms of this light curve is $10^{-3}$ an order of
magnitude higher than the input noise parameter is.

\subsection{Light curve analysis 
in our framework}

When we apply the approach of this work 
we need to calculate only one Fourier spectrum and
a single non-linear fit for each light curve.
The Fourier spectrum and the characteristics of
the light curve help us to choose the proper
fitting formula and to determine the initial
values of the fit. 

The Fourier spectrum for any of the light 
curves A and B provide us with the necessary parameters 
($f_0$, $j$, $a_0$, $a_j$, $\varphi_j$) of the carrier wave.
The amplitude modulation and its non-sinusoidal nature
in both light curves are apparent.
Searching for peaks in the low frequency range 
of the Fourier spectra results in
good initial values for $f_{\mathrm m}$, $p'=1, 2$, $a^{\mathrm A}_{p'}$
and $\varphi^{\mathrm A}_{p'}$.
While more side peaks are detectable around the harmonics of the
main frequency than the number of harmonics 
of $f_{\mathrm m}$ and the side peaks are asymmetrical, an FM
has to be assumed.  To check its non-sinusoidal nature we may
prepare e.g. an O$-$C or maximum brightness vs. maximum phase diagram.
At the end of this preparatory work we can choose the
fitting formula (\ref{mod_nsinC}). We note that
to determine the correct initial values of $a^{\mathrm 
F}_p$ and $\varphi^{\mathrm F}_p$ depends on the
tool used for finding the frequency variations.
In the worst case, they can be estimated by some
numerical trials with the non-linear fit.
Using initial values that are good enough 
the non-linear fit converged fast for models A and B. 
The algorithm reached the noise levels $10^{-2}$ and $10^{-4}$ 
within few (less than 10) iterations automatically. The residual
light curves and their Fourier spectra are 
plotted in the bottom panels of Fig~\ref{appl_Fig}.
The number of fitted parameters for both models is only 33.
Due to the finite numerical accuracy 
the residual spectra always show a structure
reflecting the original spectra at very low levels.

In conclusion our method fits the light curves
in a single step with much less parameters than the traditional one.
In addition we avoid the time-consuming alias fitting and subtracting
processes. The difference in the number of used parameters increases
with the increasing accuracy of the observed data sets.
In our example the number of regressed parameters is reduced from 
6 times (model A) to more than 10 times (model B).
Our description has advantages in the numerical fit
of the ground-based observations as well, but its advantages
are outstanding in the analysis of the space-born time series.

\section{Discussion and Summary}

\begin{table*}
 \centering
  \caption{A schematic table for classifying different 
types of single frequency modulations from
their light curves and Fourier spectra. The indices
$l$ and $k$ are integers, where $l$ is a finite number,
while $k$ can generally be infinite.}\label{sema}
  \begin{tabular}{llccc}
  \hline
modulation  &	phenomenon & AM	&	FM	& Combined \\
 \hline
 sinusoidal & 	amplitude variations:& yes (simple) &	no 	& yes (simple)  \\  
      & 	phase variations: & no    &	sinusoidal 	&  sinusoidal \\ 
      & side peak structure:	& triplets  ($jf_0\pm f_{\mathrm m}$) &	multiplets ($jf_0\pm k f_{\mathrm m}$)	& 
multiplets	($jf_0\pm k f_{\mathrm m}$) \\
      & side peak amplitudes:	& symmetrical &	symmetrical     &asymmetrical	 \\
      & 	modulation frequency: & $f_{\mathrm m}$ &  -- &  $f_{\mathrm m}$	 \\
\hline
 non-sinusoidal & 	amplitude variations:&  yes (complex)  & no 	&  yes (complex) \\ 
      & 	phase variations: & no    &	non-sinusoidal 	&  non-sinusoidal \\ 
      & side peak structure: &multiplets ($jf_0\pm lf_{\mathrm m}$) &	multiplets ($jf_0\pm k f_{\mathrm m}$)	&multiplets ($jf_0\pm k f_{\mathrm m}$)	    \\
      & side peak amplitude: &symmetrical &	asymmetrical &asymmetrical	    \\
      & 	modulation frequencies:&  $l f_{\mathrm m}$  &	 --	
& $l f_{\mathrm m}$	 \\
\hline
\end{tabular}
\end{table*}

In this paper we have investigated mathematical representation of 
artificial light curves. These light curves are defined as  
modulated signals where their carrier wave is a monoperiodic 
RR\,Lyrae light curve defined by its finite Fourier sum.
Different types of periodic functions are taken into 
account as  modulation functions from the simple sinusoidal
to multiperiodic and general ones. The consequences of these
modulation functions and modulation 
types (AM, FM, combined) are reviewed and presented.  

We followed a step-by-step analysis from the simplest case to the more 
complicated ones. The results of this process can be summarized
as follows:
\begin{itemize}
\item{(i) Tuning AM by the used  modulation function, namely
the number of harmonic terms and their amplitudes and phases,
the {\it synthetic light curves} reproduce well
the observed shape of the Blazhko envelope curves. 
This is always true for the envelopes of maxima, 
however, the envelope of minima are affected by 
the fine structure of the bump and its phase shift along the light curve 
and changing shapes. These variations in the shock are associated to the 
Blazhko modulation such a way that  
a displacement of the shock forming region occurs 
over the Blazhko cycle. This induces an earlier 
or later occurrence of the bump in the light curves 
(see \citealt{Pre65}). If these variations
are strong enough, the envelope curve of minima 
could be different from our simple modulated ones.
(ii) We showed that AM with extremely high modulation depth ($h>1$)
might be an explanation for the strange light curve shape
of V445\,Lyrae observed by {\it Kepler\/}. 
(iii) When the AM function depends on more than one frequency,
(in a parallel or cascade modulation)
we can generate envelope curves with different
observed phenomena such as beating effect, alternating maxima,
long-term, periodic amplitude changes.  
}
\item{{\it Fourier spectra of AM cases} can be easily
classified (see also Table~\ref{sema}).
For a periodic modulation the signal represented
by a finite Fourier sum the spectrum of the synthetic
light curve shows the modulation frequency,
multiplets around the pulsation frequency and its harmonics, as well.
The order of multiplets (the number of
peaks on one side) is the same as the used number of
terms for describing the modulation function. 
As a special case, the sinusoidal modulation results in
triplet structures and the appearance of $f_{\mathrm m}$ only.
 For a multiperiodic parallel modulation, the spectrum consists
of a sum of the frequencies of each component modulation,
while in the case of modulated modulation (AM cascade)
spectra includes additional peaks  
at the all possible linear combination frequencies, as well.
}
\item{
For all AM cases the Fourier {\it amplitude of the side frequencies} 
are proportional to the given harmonics' amplitude, therefore,
the numbers and the side peak amplitude ratios 
 compared to the central frequencies are the same 
for all orders of harmonics. Pure AM results in  
symmetrical multiplets: side
frequencies of the same order have the same amplitudes. 
}
\item{
The {\it variation of the mean brightness} through
the Blazhko cycle is a consequence of the AM modulation.
The effect appears even for the simplest sinusoidal
modulation function, and is closely related to
the appearance of the 
modulation frequency itself in the Fourier spectrum.
}
\item{Pure {\it FM light curves} have no amplitude changes
but significant phase variations. Their Fourier spectra
show multiplet structure around the pulsation
frequency and its harmonics even for the simplest
sinusoidal modulation. As opposed to AM
the detectable number of side peaks is increasing with the
increasing harmonic orders. The modulation frequencies
(and their harmonics) are absent in the spectra.
The side peaks' structure can be symmetrical
(for sinusoidal modulation) or asymmetrical 
(most of the non-sinusoidal cases).
The non-sinusoidal FM can be characterized by the classical O$-$C tool. 
The multifrequency FM spectra include all possible linear 
combination frequencies in the parallel case. 
}
\item{
The {\it Fourier amplitude ratio of the side peaks vs. harmonic orders} 
are determined by the Bessel functions of the first kind
(for the sinusoidal case) or their product 
(for the non-sinusoidal case).
As a consequence, these relationships are generally non-strictly 
decreasing with the increasing harmonic orders. 
Therefore, we can find larger amplitudes for the higher-order side
peaks than for the lower-orders ones for comparing the same order of side 
peaks around different harmonics and different order of side peaks 
around a given harmonic. The 
latter can be true also for the amplitude of the side
peaks and the central peaks' (the harmonics') amplitude.
}
\item{
The simultaneous AM and FM show all the above mentioned
effects. The side peaks' amplitudes in the 
Fourier spectra of the combined modulations
are generally asymmetrical already for sinusoidal AM and FM. 
The asymmetry can totally remove side peak(s) in one side.
The strength of asymmetry depends on the initial phase difference of 
the AM and FM modulation functions. This phase difference
determines both the inclination and direction of motion
of the loop in the maximum brightness vs. maximum phase diagram.  
}
\item{
In the case of the pure sinusoidal AM and FM
(and with very special parameters for non-sinusoidal and 
combined cases as well),
there are well-defined phases where the modulated and
non-modulated light curves are identical. However,
we have pointed out that, when a star shows general combined
modulation (even a sinusoidal one), 
its light curve is always different
 from any monoperiodic (non-modulated) stars.
}
\item{
We showed that our modulation description presented here
needs much (typically 3--10 times) less 
parameters for fitting a light curve than the classical solution,
where all detected frequencies have their own independent
amplitude and phase. 
The prize to pay for the relatively small
number of parameters is the more complicated fitting formulae. 
Our approach fits the light curves only once,
instead of the traditional successive prewhitening process, and
unaffected by  most aliasing problems.
}
\end{itemize}

It is equally important to list those observed features
that do not follow this picture. 
One of these effects is the {\it phase lag},
a difference between the Blazhko phases 
of maxima of maxima and minima of minima 
of the modulated light curves. These phase lags can be
several pulsation cycles long. In our study, however,
 all envelopes are highly symmetrical: their maxima and minima
are practically at the same Blazhko phases. 
If we see the well-sampled observed light curves 
(e.g. {\it CoRoT\/} or {\it Kepler\/} data) showing this 
phase lag, we can realise that the effect builds up
from the systematic motions and changing the shape 
of the bump caused by the hydrodynamical
shock in the RR\,Lyrae atmospheres.

An additional unsolved problem is 
the {\it distribution of the side peaks' asymmetry}.
The large surveys (OGLE, MACHO) found that the
spectra of Blazhko RR\,Lyrae stars more frequently contain 
asymmetrical side frequency patterns, where
the higher frequency side peaks have the higher
amplitude than vice versa 
(e.g. 74 vs. 26 per-cents for LMC by \citealt{Alc00}).
All formulae suggest 50-50 per-cents of probabilities,
if the distribution of the relative phases between AM and FM is uniform. 
The detected asymmetrical distribution might have deeper physical 
origin.  

Our description does not explain the
{\it additional frequencies} such as half-integer frequencies
belonging to period doubling, temporary overtone frequencies and
further exotic frequencies in the Fourier 
spectra, recently discovered from the space data 
\citep{Gru07, Kol10, Sza10, Cha10, Por10, Ben10}. 
Their connection with the physics of
Blazhko effect is poorly understood 
and their modelling is out of the scope of this paper.

The main result of this analysis is the demonstration  that 
the modulation paradigm gives successfully accounts for 
most of the observed properties of Blazhko RR\,Lyrae light curves.
Moreover, this work may help to distinguish between those features that
have deeper physical origin from those ones which appear
simply due to the modulation. This framework seems to
be very flexible and can easily be generalised to a
larger fraction of variable stars taking into account, 
e.g., multifrequency carrier waves or stochastic modulations.

\section*{Acknowledgments}

This project has been partially 
supported by the ESA PECS projects No.~98022 \& 98114, 
and the Hungarian OTKA grant K83790.
R.Sz. thanks the support of the J\'anos Bolyai 
Research Scholarship of the HAS. The authors thank to the
referee Dr. Katrien Kolenberg for her helpful comments and close reading 
of the manuscript.

\appendix

\section{About the exact Fourier transforms}\label{AAF}

There are many definitions of the Fourier transform,
differing from each other by their normalizations.
In this paper we adhere to the following definition:
\begin{equation}
\mathcal{F}( f )= 
\int_{-\infty}^{\infty} g(t) {\mathrm e}^{-{\mathrm i}2\pi f t} dt. 
\end{equation}
Since we always transform our equations describing
the different types of modulated signals to linear
combinations of sinusoidal functions, it is
necessary to know the exact Fourier transformation of 
these elements.  

The Fourier transform of the sinusoidal 
carrier wave Eq.~(\ref{car}) itself is
\begin{multline}\label{Fr_sinus}
\mathcal{F}\left[ c(t) \right] = 
\pi \sqrt{2\pi} U_{\mathrm c} \left\{ {\mathrm i}\cos \varphi_{\mathrm c} 
\left[ \delta(f - f_{\mathrm c}) - \delta(f + f_{\mathrm c}) \right] \right. +{}\\ 
\left. \sin \varphi_{\mathrm c} \left[ 
\delta(f - f_{\mathrm c}) + \delta(f + f_{\mathrm c}) \right] \right\}, 
\end{multline}
where i is the imaginary unit, $\delta$ is the Dirac function. 
As we see, the Fourier amplitude spectrum belonging to $c(t)$
has two components: one at positive frequency  (centred 
on $+f_{\mathrm c}$) and one at negative frequency (centred on 
$-f_{\mathrm c}$). We concern only the positive frequency, since
the negative ones have no physical meaning. 

Demonstrating the use of the above formulae,
the positive part of the Fourier transformation 
of the sinusoidal AM signal (\ref{AM1}) is given:
\begin{multline}\label{sin_AM_F}
\mathcal{F^+}\left[ U_{\mathrm {AM}}(t) \right]=
\mathcal{F^+}\left[ c(t) \right] +{} \\
\pi \sqrt{\frac{\pi}{2}} U_{\mathrm m} \left[
\left[ \cos (\varphi_{\mathrm c} - \varphi_{\mathrm m}) + 
{\mathrm i}\sin (\varphi_{\mathrm c} -\varphi_{\mathrm m}) \right] 
\delta (f+f_{\mathrm c}-f_{\mathrm m}) -{} \right. \\ 
\left. \left[ \cos (\varphi_{\mathrm c} + \varphi_{\mathrm m}) + 
{\mathrm i}\sin (\varphi_{\mathrm c} +\varphi_{\mathrm m}) \right] 
\delta (f+f_{\mathrm c}+f_{\mathrm m}) \right].   
\end{multline}
According to this schema,
the investigated signals describing the more complicated expressions 
can also be calculated in a straightforward way.

\section{Generalized product-to-sum formulae}\label{AP-S}

Let us calculate the generalised form of the
product-to-sum expression from the well-known
identity of trigonometry:
\begin{equation}\label{p-s}
\sin (\alpha_1) \sin (\alpha_2) = \frac{1}{2} 
[ \cos (\alpha_1 - \alpha_2) - \cos (\alpha_1 + \alpha_2)]. 
\end{equation}
Assuming $n$ factor in the left-hand-side it can be formally
written as $\prod_{i=1}^n\sin {\alpha_i}$.
Introducing the vector-scalar function
$\mathcal{S}_n(\boldsymbol{\alpha}):=\prod_{i=1}^n\sin 
{\alpha_i}$,
$\boldsymbol{\alpha} = (\alpha_1, \alpha_2, \dots, \alpha_n)^{\mathrm 
T}$, where T indicates transposition,
and applying the formula (\ref{p-s}) $n$ times recursively we arrive to
\begin{multline}
\mathcal{S}_n(\boldsymbol{\alpha}):=
\begin{cases}
2^{(1-n)}\sum\limits_{l=1}^{2^{(n-1)}} \left( -1 \right)^{N} 
\cos\left( \sum\limits_{i=1}^{n} Q_{li} \alpha_i \right) 
&\text{if $n$ even,}\\
\\
 2^{(1-n)}\sum\limits_{l=1}^{2^{(n-1)}} \left( -1 \right)^{N'}
\sin\left( \sum\limits_{i=1}^{n} Q_{li} \alpha_i \right) 
&\text{if $n$ odd,}
\end{cases} 
\end{multline}
where 
\begin{equation*}
N=\frac{3n}{2} - \sum_{i=1}^{n}Q_{li}, \ \ \ \text{and} \ \ \ 
N'=\frac{3(n-1)}{2} - \sum_{i=1}^{n}Q_{li}.
\end{equation*} 
Each $\sin$/$\cos$ term includes a sum of $n$ terms of the
$\alpha_i$ angles inside their arguments as
$\alpha_1\pm \alpha_2\pm \alpha_3,\dots \pm \alpha_n$.
One combination from these sets contains $l$ positive 
and $n-l$ negative angles. The total number of such a 
combination is $2^{n-1}$.
Each row of the matrix $Q$ contains the signs 
of the angles of a possible set from the total number of $2^{n-1}$.
Namely, the elements of the matrix $Q$ are either
 $+1$ or $-1$, but $Q_{l1}=1$ i.e. 
the first columns contain $+1$ for all $l$s.

\bsp

\label{lastpage}

\end{document}